\documentclass[usenatbib, useAMS]{mnras}
\usepackage{amsmath}
\usepackage{bm}
\usepackage{newtxtext,newtxmath}
\usepackage{color}
\usepackage{tensor}
\usepackage{multirow}
\usepackage{graphicx}
\usepackage{natbib}
\usepackage{myaasmacros}
\usepackage{enumerate}
\usepackage[english]{babel}
\usepackage{tabularx}
\usepackage{placeins}

\newcommand{\Msol}{\mathrm{M}_{\odot}}
\newcommand{\kpc}{\mathrm{kpc}}
\newcommand{\Mpc}{\mathrm{Mpc}}
\newcommand{\Gpc}{\mathrm{Gpc}}

\newcommand{\s}{\mathrm{s}}
\newcommand{\km}{\mathrm{km}}
\newcommand{\kmps}{\km\,\s^{-1}}

\def\lsim{\mathrel{\rlap{\lower3pt\hbox{$\sim$}}
    \raise1pt\hbox{$<$}}}                % comment removed
\def\gsim{\mathrel{\rlap{\lower3pt\hbox{$\sim$}}
    \raise1pt\hbox{$>$}}}                % comment removed

\begin{document}
\pubyear{2020}
\title[Suppressing numerical diffusion]{EDGE: A new approach to suppressing numerical diffusion in \\ adaptive mesh simulations of galaxy formation}
\author[A. Pontzen et al]{Andrew Pontzen$^1$\thanks{a.pontzen@ucl.ac.uk}, 
Martin P. Rey$^{2,1}$, Corentin Cadiou$^1$, 
Oscar Agertz$^2$,\newauthor  Romain Teyssier$^3$, 
Justin Read$^4$, Matthew D. A. Orkney$^4$
  \\
  $^1$ {Department of Physics and Astronomy, University College London, London WC1E 6BT} \\
  $^2$ {Lund Observatory, Department of Astronomy and Theoretical Physics, Lund University, Box 43, SE-221 00, Lund, Sweden} \\
  $^3$ {Institute for Computational Science, University of Zurich, Winterthurerstrasse 190, 8057 Zurich, Switzerland} \\
  $^4$ {Department of Physics, University of Surrey, Guildford GU2 7XH, UK} \\
  }

\date{ Received ---; published---. }
\maketitle

\newcommand{\pc}{\mathrm{pc}}
\newcommand{\cm}{\mathrm{cm}}
\newcommand{\Myr}{\mathrm{Myr}}
\newcommand{\dd}{\mathrm{d}}
\newcommand{\Gyr}{\mathrm{Gyr}}
\newcommand{\ergs}{\mathrm{ergs}}
\renewcommand{\vec}[1]{\bm{#1}}
\newcommand{\ee}{\mathrm{e}}

\begin{abstract}
  We introduce a new method to mitigate numerical diffusion in adaptive mesh refinement (AMR) simulations of cosmological galaxy formation, and study its impact on a simulated dwarf galaxy as part of the `EDGE' project. The target galaxy has a maximum circular velocity of $21\,\kmps$ but evolves in a region which is moving at up to $90\,\kmps$ relative to the hydrodynamic grid. In the absence of any mitigation, diffusion softens the filaments feeding our galaxy. As a result, gas is unphysically held in the circumgalactic medium around the galaxy for $320\,\Myr$, delaying the onset of star formation until cooling and collapse eventually triggers an initial starburst at $z=9$. Using genetic modification, we produce `velocity-zeroed' initial conditions in which the grid-relative streaming is strongly suppressed; by design, the change does not significantly modify the large scale structure or dark matter accretion history. The resulting simulation recovers a more physical, gradual onset of star formation starting at $z=17$. While the final stellar masses are nearly consistent ($4.8 \times 10^6\,\Msol$ and $4.4\times 10^6\,\Msol$ for unmodified and velocity-zeroed respectively), the dynamical and morphological structure of the $z=0$ dwarf galaxies are markedly different due to the contrasting histories. 
  Our approach to diffusion suppression is suitable for any AMR zoom cosmological galaxy formation simulations, and is especially recommended for those of small galaxies at high redshift.
\end{abstract}


\section{Introduction}
Numerical hydrodynamics plays a central role in understanding the formation and evolution of galaxies, and their impact on the Universe \citep[e.g.][]{SomervilleDave15}. However, numerical methods for solving the hydrodynamical equations are required to trade off accuracy in different regimes. The two most established methods, smoothed particle hydrodynamics (SPH) and adaptive mesh refinement (AMR) have well-documented, complementary strengths and weaknesses \citep{Agertz07,Springel10SPHReview,Teyssier15AMRReview}. While the `subgrid' physics (i.e. differing approaches to incorporating unresolved astrophysical processes) dominates the uncertainty budget of galaxy formation simulations \citep{Aquila12,KimAgora16}, errors in hydrodynamics  can also cause systematic shifts in galactic properties, so should be studied and suppressed to the greatest possible degree.  

One approach is to combine the best of SPH and AMR worlds by using a moving mesh \citep[e.g.][]{Arepo10,Arepo19} or other hybrid solver \citep{Hopkins15}. On the other hand, the comparative immaturity of the resulting numerical integrators means that their properties are still relatively unexplored. Movement in the mesh is known to exacerbate numerical noise associated with the grid itself, which can then act as a source of instabilities; the importance of this to physical solutions remains unclear \citep{McNally12,Munoz14}. There are also strong practical and sociological reasons why   groups  choose to work with codes into which they have invested significant effort and where strengths and weaknesses are familiar.

In this paper we will focus on the best-known weakness of AMR: its results are dependent on the rate at which the fluid moves relative to the grid, breaking Galilean invariance \citep[e.g.][]{Robertson10}. The effect of grid-relative motions is to cause numerical diffusion; such diffusion is in fact required to produce stable hydrodynamical solutions, so is in equal part a strength and weakness of successful schemes \citep{Teyssier15AMRReview}. In some circumstances the diffusion is negligible but if the entire region being simulated moves rapidly relative to the grid, diffusion may degrade the effective resolution to be many times worse than the grid spacing would suggest. 

In this work, we will study diffusion in AMR simulations of dwarf galaxies, and discuss a new route to mitigating its effects that we initially designed for the EDGE\footnote{Engineering Dwarfs at Galaxy formation's Edge} project \citep{2019arXiv190402723A}. The EDGE regime is particularly challenging: we target small galaxies (virial velocities $\simeq 20\,\kmps$), where the impact of numerical diffusion is most significant compared with the gravitational potential. While the {\it mean} velocity in the cosmological box is zero by construction, the galaxies typically move at $\simeq 80\,\kmps$ relative to the grid, an order of magnitude higher than the internal velocity dispersions. 

The rate of flow is only one factor in determining the extent of numerical diffusion; it is also sensitive to the order of the hydrodynamic solver, the cell size, and the timestep \citep[e.g.][]{Robertson10}. 
Since AMR demonstrably converges in the limit of infinitely small cells, the simplest route to minimising diffusion is to use an aggressive refinement scheme \citep[e.g.][]{Few16}; however, as we will see in this work, such an approach is computationally expensive in general. While most refinement schemes are quasi-Lagrangian, i.e. aiming to achieve roughly equal mass per grid cell, using a super-Lagrangian scheme \citep[e.g.][]{Chabanier20} is one route to suppressing numerical diffusion in a more targeted (and therefore computationally cheaper) region. 
A separate option is to use higher-order slope estimation schemes; however, according to Godunov's theorem, only first order linear schemes are monotonicity preserving, meaning higher-order solutions are oscillatory and may generate negative densities. Slope limiters, which smoothly reduce the order of the solver in regions with steep gradients, can be used to mitigate this problem at the expense of making the method more diffusive. There are many different possible limiters, each trading stability and diffusion in a different way \citep[e.g.][]{Toro99}. 

High-order slope estimation would be particularly challenging in our galaxy formation simulations. We model stellar feedback by directly modifying gas properties in cells surrounding young star particles, often resulting in sharp discontinuities within the density, velocity and energy fields.  We have chosen the MinMod slope limiter \citep[e.g.][]{Toro99} which is the safest choice in terms of suppressing instabilities, but consequently also the most diffusive. We find that, if we adopt a less diffusive slope limiter, the solution becomes unphysical; with the current implementation of feedback, we are hence required to adopt the MinMod slope limiter in order to maintain stability. In contrast, galaxy simulations accounting for strong stellar feedback via source terms in the Riemann solver, e.g. using a non-thermal pressure \citep{Agertz13,Kretschmer2020}, have successfully been carried out using slope limiters such as MonCen \citep{AgertzKravtsov2015}. We leave a more systematic investigation of this topic for future work, and instead focus here on a completely new solution to the diffusion problem.

If neither the solver nor the cell size can easily be changed, the final alternative is to ensure the rate of flow relative to the grid is small. However, halos with a naturally small net velocity are exceptionally rare. For example, in the region used by the EDGE project, $95\,\%$ of all halos move at velocities greater than $40\,\kmps$ at the time of early star formation, $z=16$. As we will see, this is unavoidable in cosmological volumes, and already enough to cause substantive numerical errors. Further, adding a spatially uniform velocity offset to the initial conditions cannot usefully reduce even a single halo's net speed: the cosmological expansion `redshifts' the momentum of particles. Any uniform velocity offset that is not associated with a potential gradient therefore decays proportionally to $1+z$; in contrast, the physical velocities grow (during matter domination in the linear regime) as $(1+z)^{-1}$. Therefore, uniform velocity offsets rapidly become insignificant compared to the growing physical velocities.

Our solution is to use the genetic modification approach \citep{Roth:2015wha,Rey18,CodePaper} to post-hoc modify the initial conditions such that the velocity of a chosen halo is minimised while its environment, accretion history and other properties are minimally affected. In fact we have already adopted this approach as a matter of course for published EDGE results \citep{Rey19,Rey20}. In this paper we examine the effects of {\it disabling} the modification in one of our galaxies in order to compare the AMR solution with and without rapid streaming. 

In Section~\ref{sec:sims} we describe the simulation and modification techniques. Section~\ref{sec:results} presents the comparison between simulated galaxies with and without the velocity-zeroing in the initial conditions. We summarise and provide recommendations for future investigations in Section~\ref{sec:disc}, taking into account the speeds at which galaxies stream in typical cosmological simulations.  Appendix~\ref{sec:advection-test} gives details of an estimate for the scale of numerical diffusion in our simulations, while Appendix~\ref{sec:vel-calc} calculates the expected streaming speeds in cosmological simulations as a function of box size.

\begin{figure}
  \includegraphics[width=\columnwidth]{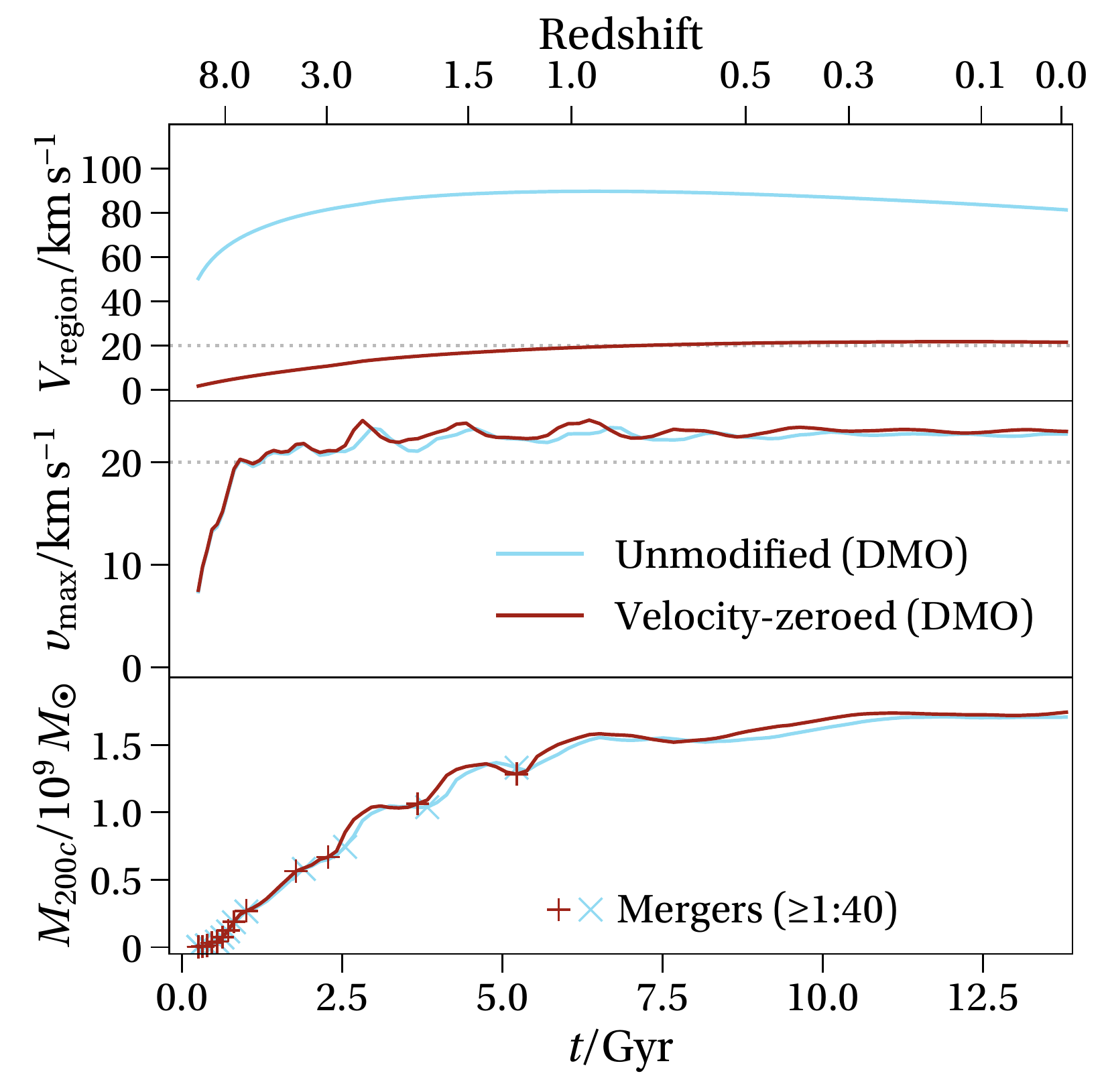}
  \caption{Lagrangian region streaming velocities relative to the simulation grid (top) contrasted with the maximum circular velocities (middle) for dark-matter-only simulations of the EDGE galaxy studied in this work. The horizontal dotted gray line on each panel shows $20\,\kmps$ for reference.  
  The mean circular velocity of $21\,\kmps$ is more than four times smaller than the peak region streaming velocity of $90\,\kmps$ in the original run (dotted lines), raising our initial concern that numerical diffusion may have a significant effect on this object. Once our modification is activated, the region velocity is zeroed in the linear initial conditions. This leaves only a higher-order residual drift which averages $11\,\kmps$; it is even smaller at high redshift where the effects of diffusion are most important.
  The accretion history (bottom) remains unchanged -- here we show, for both simulations, the virial mass $M_{200c}$ as a function of time; we also mark the mergers with ratios at least 1:40 with crosses. The results from the two initial conditions are almost indistinguishable, because the modification to the velocity is by construction highly coherent across the scales from which the halo forms (Figure~\ref{fig:ICs}). }\label{fig:DMO-history}
  \end{figure}

\begin{figure*}
  \includegraphics[width=1.0\textwidth]{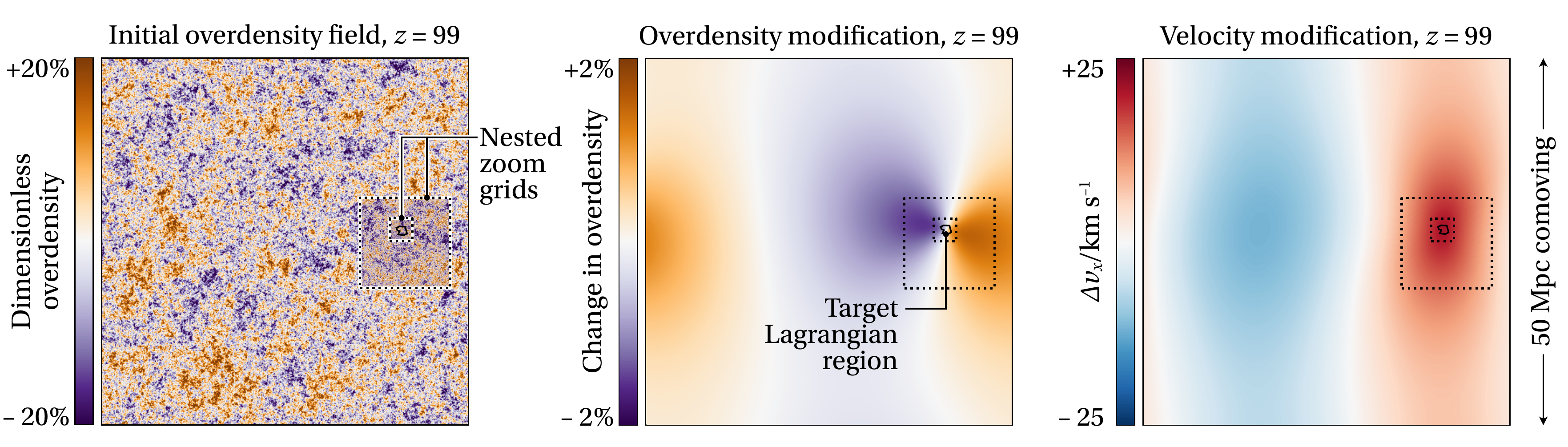}
  \caption{A slice through the overdensity field in our initial conditions at $z=99$ (left panel) and the genetic modification that we apply to make the linear-theory velocity of the Lagrangian region zero (centre panel, colour scale exaggerated by a factor ten relative to the left panel). The $x$-component of the resulting change in the linear velocity field is shown in the right panel. This illustrates how, to effect a change in the velocity, {\sc genetIC} creates gradients in the potential by marginally altering very long-wavelength modes in the overdensity. These have negligible effect on the local large scale structure from which the galaxy forms (see Figure~\ref{fig:structure-panels}) because the Lagrangian region is tiny compared to the correlation length of the velocity. All panels show the location of our nested zoom grids which are required to achieve an effective resolution of $32\,768^3$.  }\label{fig:ICs}
\end{figure*}

\section{Galaxy formation simulations}\label{sec:sims}

Initial conditions are generated using the \textsc{genetIC} code  which permits modifications to be made to a Gaussian random field realised at increasing resolution in nested zoom boxes \citep{CodePaper}. The parent box has $512^3$ particles across a box length of $L=50\,\mathrm{Mpc}$ comoving and is generated to match the cosmological parameters $\Omega_M = 0.309$; $\Omega_\Lambda=0.691$; $\Omega_b = 0.045$; $H_0 = 67.77 \,\mathrm{km\,s^{-1}\,Mpc^{-1}}$ \citep{2014A&A...571A..16P}. Two zoom grids are opened with side lengths $12.5$ and $6.25\,\mathrm{Mpc}$ respectively, with the deepest level attaining a resolution equivalent to $32\,768^3$ particles on the base grid. 

The zooms are selected in two stages. First, we simulate the uniform-resolution box  with dark matter only, and visually identify a void which we resimulate at the equivalent of $2048^3$ resolution. We then identify all dark matter halos within the void at $z=0$ using HOP \citep{1998ApJ...498..137E} and search for candidate dwarf galaxies with no massive neighbours and a suitable virial mass $10^9 < M_{200c} / M_{\odot} < 10^{10}$. Here, $M_{200c}$ is defined as the mass inside a spherical volume encompassing 200 times the cosmic critical density and we will also make use of the corresponding virial radius, $r_{200c}$. We consider only isolated objects, by which we mean those without an equally or more massive counterpart within $10$ virial radii at $z=0$; for more details see \cite{2019arXiv190402723A}. In this work, we will focus on a single example with $M_{200c} \simeq  1.6 \times 10^{9} M_{\odot}$, which is the `Reference' galaxy from \cite{Rey19}. 

We perform an initial dark-matter-only zoomed simulation of our object. All simulations are performed using \textsc{Ramses} \citep{Teyssier02Ramses} and analysed with the combination of \textsc{pynbody} \citep{2013ascl.soft05002P} and \textsc{tangos} \citep{2018ApJS..237...23P}. At each output, we measure the halo circular velocity $v_{\mathrm{max}}$, defined as the maximum of $v_c$ where $v_c^2 = GM(<r)/r$, and $M(<r)$ is the mass enclosed in spheres $r$ around the halo centre.  We also measure the mass-weighted mean velocity of the Lagrangian region $v_{\mathrm{region}}$, defined by all dark matter particles that will fall into the halo by $z=0$.

The top two panels of Figure~\ref{fig:DMO-history} show these quantities as a function of time. Blue lines denote the results from the original initial conditions; we find that the region streams at $51\,\kmps$ at $z=16$ (a time at which we will find our hydrodynamical simulations start to form stars) and grows to a peak of $90 \mathrm{\,km\,s^{-1}}$ at $z=1$, compared to a halo maximum circular velocity of $21\,\mathrm{km\,s^{-1}}$. The halo's potential well is therefore shallow compared to the flow rate across the grid.

We now discuss how to suppress the streaming in the initial conditions. A slice through the unmodified zoom initial conditions for the chosen object is shown in Figure~\ref{fig:ICs}.  In the linear regime the velocity field is uniquely determined by the overdensity field (through the continuity equation). Consequently we may zero the mean velocity in the target Lagrangian region  through a suitable change to the overdensity field. We follow the procedure described by \cite{Roth:2015wha}, which is a variant of the \cite{HoffmanRibak91} algorithm. Modifications made in this way change the field to the minimum extent necessary to obtain the desired alteration, while retaining a high likelihood within the cosmological Gaussian random ensemble.  For more information see \cite{vanDeWeygaert96} for the relation between velocity and overdensity constraints; \cite{Roth:2015wha} and \cite{Rey18} for the principles of modification and their close relationship to constraints; and \cite{CodePaper} for the details of our multigrid algorithm that implements the required modifications in practice.

The outcome of the operation is shown in the middle and right panels of Figure~\ref{fig:ICs}. The middle panel shows the change in the overdensity required to zero the linear velocity, exaggerated in colour scale by a factor ten relative to the field in the left panel. The right panel shows the change in the $x$ component of the corresponding velocity field. These show how the algorithm generates a shift in the velocity by adding a very large scale density gradient to the box.    We refer to simulations with the modified initial conditions as `velocity-zeroed', and we verified that the velocity of the Lagrangian region in the initial conditions is indeed precisely zero. 

We perform a new dark-matter-only simulation, now starting from the updated initial conditions. Non-linear evolution implies that the halo is not perfectly stationary relative to the grid at late times, but the net motion is substantially suppressed. The typical streaming velocity in the modified simulation is $\simeq 10\,\mathrm{km\,s^{-1}}$ (peaking at $22\,\mathrm{km\,s^{-1}}$ by $z=0$); see the red line in the upper panel of Figure~\ref{fig:DMO-history}. It remains smaller than the circular velocity of the halo at all times, and is close to zero at high redshift.

The velocity modifications are extremely coherent across the Lagrangian patch of the target halo meaning that there is virtually no change in the formation history of our halo or the surrounding structures.  The mass accretion histories in the original and velocity-zeroed cases are shown in the lower panel of Figure~\ref{fig:DMO-history}; they match closely. We indicate all mergers more significant than 1:40 (using $\times$ symbols for mergers in the unmodified case, and $+$ in the velocity-zeroed case), and these also show good agreement in timing between the two versions of the simulation. The left panels of Figure~\ref{fig:structure-panels} illustrate the large scale structure, showing the projected dark matter density in the zoom region at $z=15.7$; there is barely any change between the two cases. This lack of any major effects at the level of the dark matter simulation, coupled to the increased gas physics fidelity that we will describe in Section~\ref{sec:results}, has led us to adopt velocity-zeroed initial conditions as the default for other EDGE studies starting with \cite{Rey19}. 

In addition to the numerical effects of grid-relative streaming, there is also a physical effect from streaming between baryons and dark matter \citep{TseliakhovichHirata10}. This streaming can suppress the condensation of gas into small halos at high redshift, which is a significant consideration in the formation of the first stars \citep{Bromm13FirstStars} but concerns earlier times and smaller halos than we consider here. These physical relative velocities decay as the universe expands, and even at the start of our simulation ($z=100$) have an r.m.s. of  $\simeq 3\,\kmps$, far smaller than the velocities relevant to numerical diffusion; they have been neglected in the present study.

We next perform three hydrodynamical simulations with velocity-zeroed and three with unmodified initial conditions. The purpose of repeating each three times is to ensure we can distinguish the improvement from velocity-zeroing from stochastic effects that may cause intrinsic variation between runs \citep{Keller19,Genel19}. The integration for both types of initial condition is performed  using a HLLC Riemann solver \citep{1994ShWav...4...25T} and the MinMod slope limiter to construct gas variables at cell interfaces. Adaptive mesh refinement follows a quasi-Lagrangian strategy, with cells being split once the number of dark matter particles exceeds 8. Refinement is only permitted for zoom particles, and is disabled beyond {\sc Ramses} level 24, i.e. the smallest cell permitted is $2^{24}$ times smaller than the simulation length, corresponding to a cell size of $3\,\mathrm{pc}$. Because the code solves gravitational forces on the grid, this is also the smallest softening length in the refined region. 

Finally, we undertake further hydrodynamical simulations to test for the effect of improved grid resolution which, as discussed in the Introduction, is guaranteed to reduce the effects of numerical diffusion. We performed two versions of such an enhanced-resolution simulation; the first reduces the threshold for refinement to~$2$ (rather than~$8$) dark matter particles, while the second maintains the same number threshold but increases the mass resolution of the dark matter particles themselves by a factor~$8$. The two approaches yield almost identical results, and we present results only from the first. The chief problem with forcing high resolution in either of these ways is the computational cost, as shown in Table~\ref{tab:runtimes}. We were only able to continue the simulation to $z=6.6$, at which point it had consumed almost $120,000$ core hours (on $160$ cores). By contrast, the unmodified runs were able to reach $z=0$ within approximately $30,000$ core hours. Velocity zeroing the initial conditions leads to a modest speed {\it improvement} compared with using unmodified initial conditions, possibly because the decreased velocities permit longer timesteps. 

\begin{table}
  \begin{center}
\begin{tabular}{rccc}
  & $z=15.7$ & $z=6.6$ & $z=0$ \\ \hline
  Unmodified & 161 & 7,098 & 32,236 \\
  Unmodified, high resolution & 581 & 119,598 & --- \\
  Velocity zeroed, original resolution & 130 & 5,381 & 28,708 \\
  \hline
\end{tabular}
\end{center}
\caption{Required number of computational core hours for our hydrodynamic simulations to reach different redshifts. From top to bottom we report respectively for the median of three simulations with unmodified initial conditions; for the simulation with forced refinement (giving higher resolution throughout the intergalactic medium); and for the median of three simulations at the original resolution but with velocity zeroing in the initial conditions. The running total is given at $z=15.7$ (as star formation is due to commence), at $z=6.6$ (during the tail end of reionisation) and at $z=0$. The high resolution run was too slow to continue beyond $z=6.6$.}\label{tab:runtimes}
\end{table}

All simulations  implement radiative cooling, star formation, feedback, metal enrichment and the cosmic UV background following the `weak' scheme of \cite{2019arXiv190402723A}. In brief, star formation obeys a Schmidt law, and is only allowed to proceed in cold gas ($T<100\,\mathrm{K}$) with densities exceeding $300\,m_p\,\cm^{-3}$ where $m_p$ is the proton mass.  Feedback injects energy, momentum, mass and heavy elements into the interstellar medium from SNIa, SNII and stellar winds.  The energy injection scheme includes a temperature ceiling $T<10^8\,\mathrm{K}$ and velocity ceiling $v<10^3\,\mathrm{km\,s^{-1}}$; these limitations are computationally efficient but unphysical and so our results should not be directly compared  to observations. The focus of the current work is comparisons between different simulations. In particular, we will see that numerical diffusion is most pronounced in the intergalactic medium surrounding our galaxies and therefore we would expect the issues discussed in this paper to be insensitive to  uncertainties in the feedback prescription.

\section{Results}\label{sec:results}

\begin{figure*}
  \includegraphics[width=1.0\textwidth]{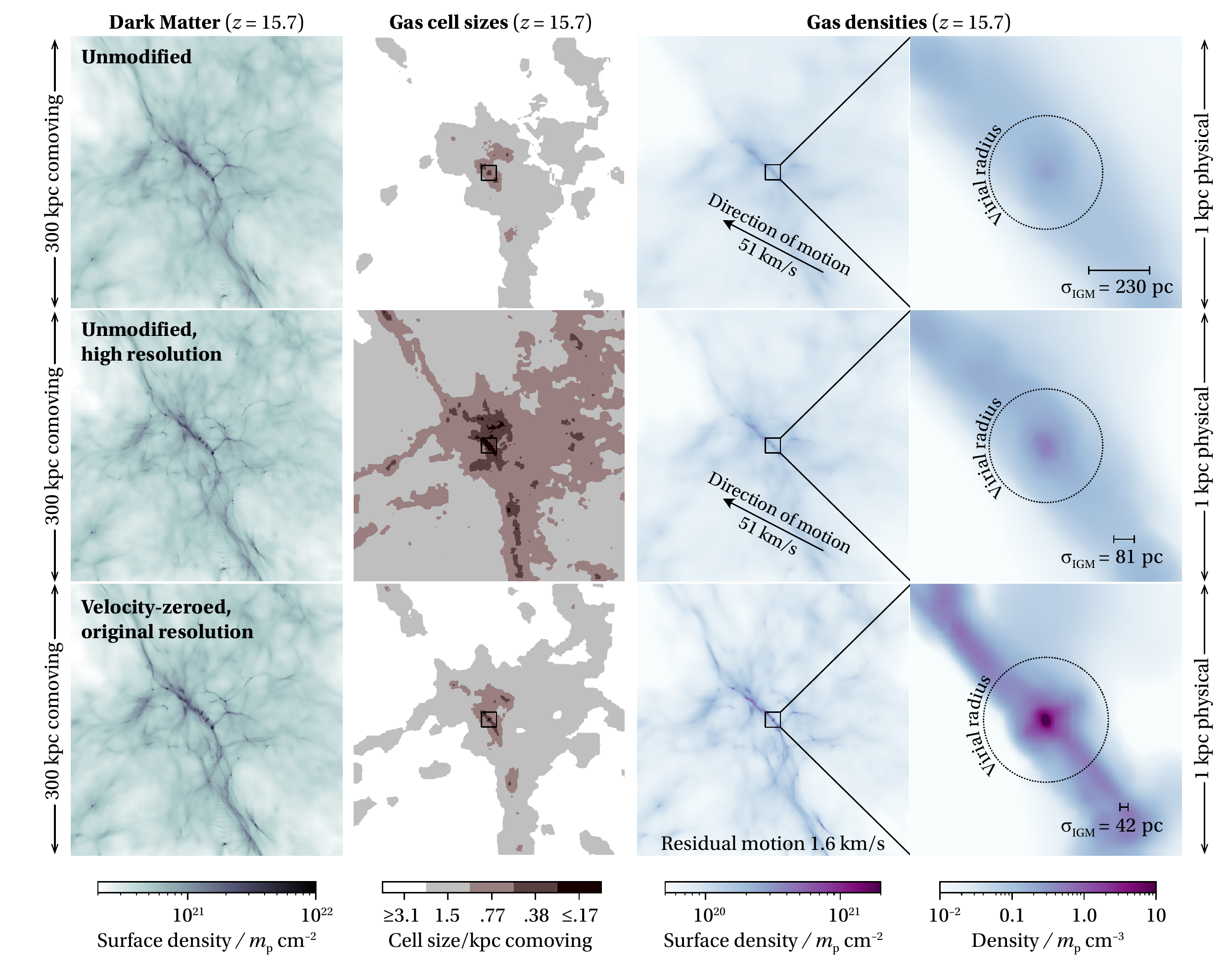}
  \caption{The effective resolution of a simulation depends on numerical diffusion, which is a function both of simulation cell size and net motion relative to the grid. Here we compare, at $z=15.7$, the effect on our simulations of starting from the original initial conditions (upper panels) with our standard refinement criterion; forcing an extra level of refinement (middle panels); and instead changing the velocity structure to perform the simulation near the galaxy's rest frame (bottom panels).  The left panels show the dark matter distribution which is nearly unaffected. This reflects the large-scale coherence of the modifications made to the velocity fields (right panel of Figure~\ref{fig:ICs}). The second set of panels from the left show the gas cell sizes, showing that the velocity-zeroed simulation has a similar refinement map to the unmodified simulation, while by construction the high-resolution simulation has much smaller cells throughout the region depicted (and so is more computationally expensive; see Table~\ref{tab:runtimes}). The third set of panels show the gas distribution, which is softened in the unmodified case due to numerical diffusion from advection. Increasing the refinement or suppressing the velocities both reduce the extent of numerical diffusion. The right panels show a zoom-in to the region in which the galaxy is forming, with the virial radius $r_{200c}$ indicated by a dotted circle. Scale bars show the estimated numerical diffusion scales in the intergalactic medium near the galaxy, $\sigma_{\mathrm{IGM}}$, which we obtain with Eq.~\eqref{eq:diffusion-scale-estimator}. }\label{fig:structure-panels}
  \end{figure*}

Figure~\ref{fig:structure-panels} shows snapshots of (from left to right) the dark matter, adaptive mesh refinement structure, gas density in the large scale structure, and a closer view of gas densities in the galaxy, all at $z=15.7$. The top panels show the unmodified simulation, the middle panels show the effect of forcing additional mesh refinement, and the bottom panels show results from adopting velocity-zeroed initial conditions\footnote{For the unmodified and velocity-zeroed cases, we show the first of the three simulations that we performed to check for stochastic effects; we verified that there is a close visual agreement between the re-runs.}. While the dark matter large scale structure agrees well between all versions of the simulation, the gas in filaments is softened to differing degrees.  Because the filament is poorly resolved when using the unmodified initial conditions, gas struggles to cool within the galaxy's shallow potential well. As a result, while star formation has commenced at $z=15.7$ in the velocity-zeroed case, it is yet to do so in the other simulations. Differences in the gas structure persist until reionisation heats the gas at $z \simeq 6$, introducing thermal pressure into the simulations and so smoothing the gas distribution on much larger scales than the filaments illustrated here.  We will first verify that numerical diffusion accounts for the differences between simulations (Section~\ref{sec:diffusion-scale}), before giving an overview of the impact on the galaxy (Section \ref{sec:gal-sfh}).

\subsection{Scale of numerical diffusion}\label{sec:diffusion-scale}
In Appendix~\ref{sec:advection-test}, we outline a 1D advection test, supported by analytic arguments, to show that the expected  scale $\sigma$ of diffusion on a fixed grid with our adopted solver is given approximately by 
\begin{equation}
  \sigma \approx 0.9 L^{0.29} \Delta^{0.71}\,,\label{eq:diffusion-scale-estimator}
\end{equation}
where $L$ is the distance travelled by the fluid across the grid and $\Delta$ is the size of a cell. The diffusion scale $\sigma$ can be reduced by decreasing one or both of $L$ and $\Delta$, in agreement with expectations.  We do not take into account refinement, cooling or self-gravity in this simple estimate, and $\sigma$ should therefore be taken only as a rough indicator of the magnitude of numerical diffusion. 

Nonetheless, we use Eq.~\ref{eq:diffusion-scale-estimator} to create a basic cross-check by estimating the scale of numerical diffusion in the intergalactic medium at $z=15.7$, when star formation is about to begin in the velocity-zeroed run. The total comoving distance travelled by the protogalactic region between $z=99$ and $z=15.7$ can be directly measured from our outputs to be $L = 286\,\mathrm{kpc}$ and $5\,\mathrm{kpc}$ in the unmodified and velocity-zeroed runs respectively. We can also study the refinement maps (Figure~\ref{fig:structure-panels}, second panels from left) to establish the cell size in the intergalactic medium filaments. From top to bottom, for the unmodified, high-resolution and velocity-zeroed runs respectively, we obtain $\Delta = 0.77, 0.17$ and $0.38\,\kpc$ comoving. With these values, we evaluate from Eq.~\eqref{eq:diffusion-scale-estimator} $\sigma_{\mathrm{IGM}}=3.9$, $1.3$ and $0.7\,$kpc comoving ($230$, $80$ and $40\,$pc physical) respectively. 

The right panels of Figure~\ref{fig:structure-panels} show scale bars with these estimated diffusion scales. The minimum scale of structure in the intergalactic medium appears in each case to agree with these approximate smoothing lengths, from which we conclude the numerical diffusion effects are indeed a dominating factor in determining the intergalactic medium densities. We verified that, once reionisation heats the intergalactic medium, the thermal pressure erases these differences. 

In the case of the velocity-zeroed simulation, part of the reduction in diffusion scale comes from mesh refinement (i.e. its $\Delta$ is smaller than that of the original unmodified run). Without including this refinement, the diffusion scale would instead be $70\,\mathrm{pc}$ physical, which is still a major improvement on the unmodified simulation ($230\,\mathrm{pc}$).
The modest initial reduction in diffusion allows for gravity to produce densities which are sufficient to trigger cooling and, in turn, refinement. Inside the galaxy, this virtuous cycle continues, leading to an even smaller diffusion scale.   The central cooling time of the velocity-zeroed simulation is $10\,\Myr$, compared to $300\,\Myr$ for the unmodified run and $200\,\Myr$ for the high-resolution case. Provided the cooling timescale becomes short compared to the age of the Universe ($250\,\Myr$ at $z=15.7$), diffusion is no longer a barrier to star formation.   

We can conclude that our velocity-zeroing approach has suppressed numerical errors to the point where they no longer hold up high-redshift star formation. We will now confirm this directly by measuring the star formation rates in the simulations.

\subsection{Effects of numerical diffusion}\label{sec:gal-sfh}

\begin{figure}
  \includegraphics[width=\columnwidth]{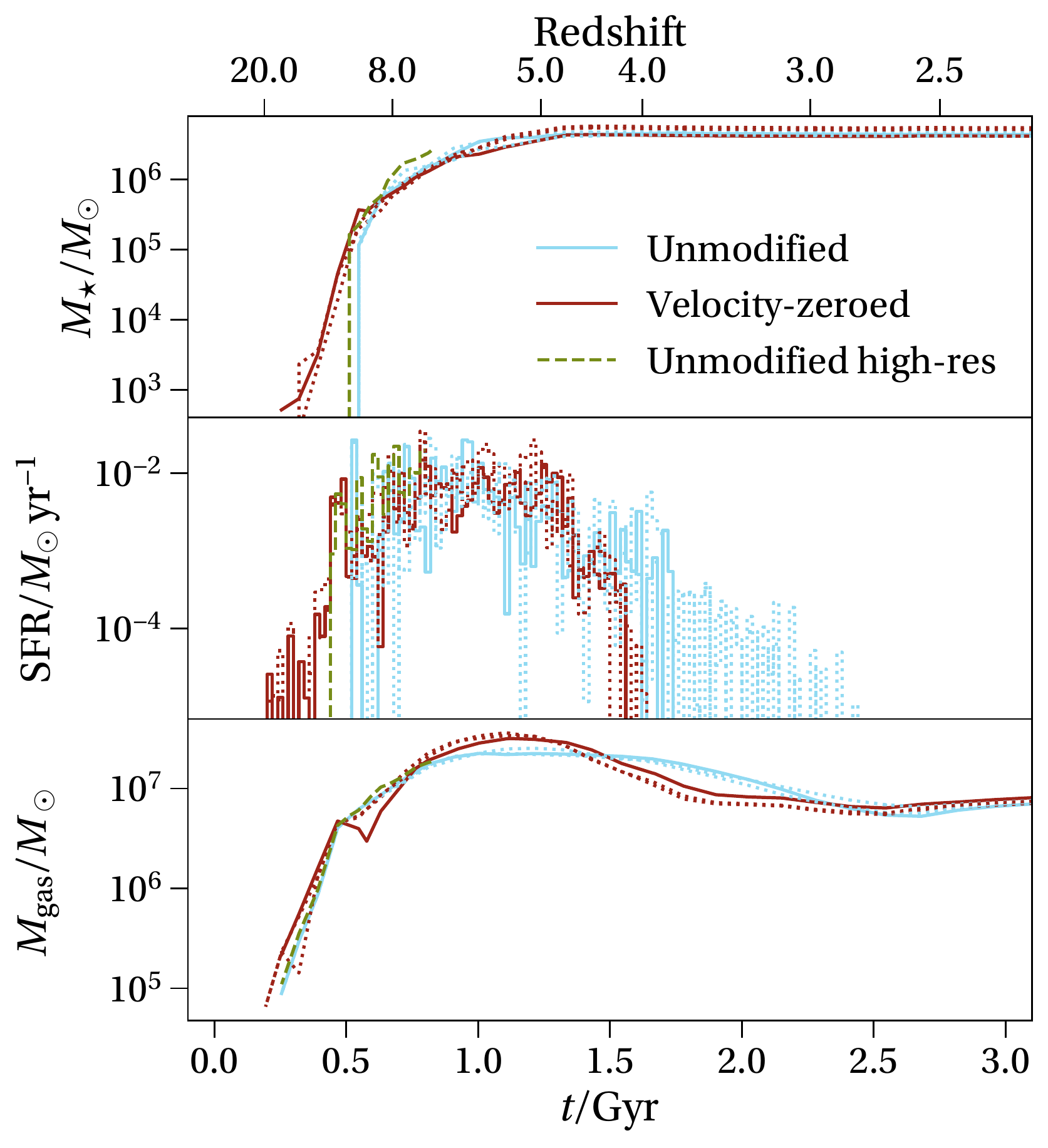}
  \caption{ From top to bottom, the total mass of stars in the galaxy,  the star formation rate, and the gas mass interior to the virial radius $r_{200c}$ as a function of time for unmodified (blue) and velocity-zeroed
    (red) initial conditions. Dotted lines show reruns to check for stochastic effects, while the dashed green line shows results from the  high-resolution simulation (which terminates at $z=6.6$). The most physical outcome is obtained by using velocity-zeroed initial conditions, with which star formation starts gradually at $z \simeq 16$. The unmodified simulations, which have rapid streaming relative to the grid, delay star formation by  $320\,\mathrm{Myr}$ and $240\,\Myr$ (for standard and high resolution respectively) due to numerical diffusion. They then undergo a compensating burst so that the final stellar mass is comparable between all cases. 
     }\label{fig:SF-etc}
  \end{figure}

  We now contrast the properties of our dwarf galaxy simulations in the unmodified and velocity-zeroed cases. We will quote quantitative values based on the two means: of the three  hydrodynamical simulations with the unmodified initial conditions, and of the three with velocity-zeroed initial conditions. This also allows us to quote uncertainties, given by the half-width of the full spread of results from the respective runs. 

We have already established that numerical diffusion due to grid-relative streaming prevents dense gas from accumulating at early times. Figure~\ref{fig:SF-etc} illustrates the effect of this on the properties of our dwarf galaxies; from top to bottom it shows the total stellar mass, the star formation rate, and the mass of gas inside the virial radius as a function of time up to  $t=3.1\,\Gyr$ ($z=2.1$). No star formation activity takes place after this time, although the simulations run to $z=0$.

The original initial conditions (blue lines) give rise to a galaxy that starts forming stars in an intense initial burst at $z=9$, forming $1.1 \pm 0.1 \times 10^5\,\Msol$ in less than $80\,\Myr$ (the time between individual outputs).
By contrast, when zeroing the initial velocity (red), star formation ramps up gradually between $z=17$ and $z=9$, which is the expected behaviour prior to reionisation, since gravitational collapse should deliver a constant stream of gas to the centre of the proto-galaxy. The high resolution simulation (green dashed line) has a star formation history that is intermediate between these two, as expected given the extent of numerical diffusion (Figure~\ref{fig:structure-panels}). Overall, the onset of star formation is delayed by $320\,\Myr$ and $240\,\Myr$ respectively in the unmodified and high-resolution cases, but the stellar mass rapidly catches up due to the strong initial bursts. 

After $z=9$, all simulations continue to form stars for a significant period.  Even though reionisation imposes an intergalactic medium thermal floor ($>10^4\,\mathrm{K}$), our prescription for self-shielding means that residual cool gas can be retained inside the galaxy and continues forming stars. Note also that our reionisation completes relatively late, at around $z=6$, corresponding to the expected thermal history in a void \citep[see][]{Rey20}.  There is considerable scatter in the final time at which residual star formation ceases, truncating at $z=4.0\pm 0.2$ for the velocity-zeroed runs but as late as $z=3.0$ when adopting the original initial conditions. The origin of the very late star formation is a dense knot of gas which is self-shielding and resistant to disruption by feedback. As discussed in Section~\ref{sec:sims}, the feedback implemented in these simulations is somewhat weaker than the `fiducial' recipe adopted elsewhere in our suite \citep{2019arXiv190402723A}, but the effect of rapid bulk flow appears to make the feedback weaker still, perhaps because  high temperature pockets of gas diffuse instead of expanding.  We verified that such dense, persistent star forming knots are not seen in any of our galaxies, irrespective of feedback details, once velocity-zeroed initial conditions are adopted.

\begin{figure}
  \includegraphics[width=\columnwidth]{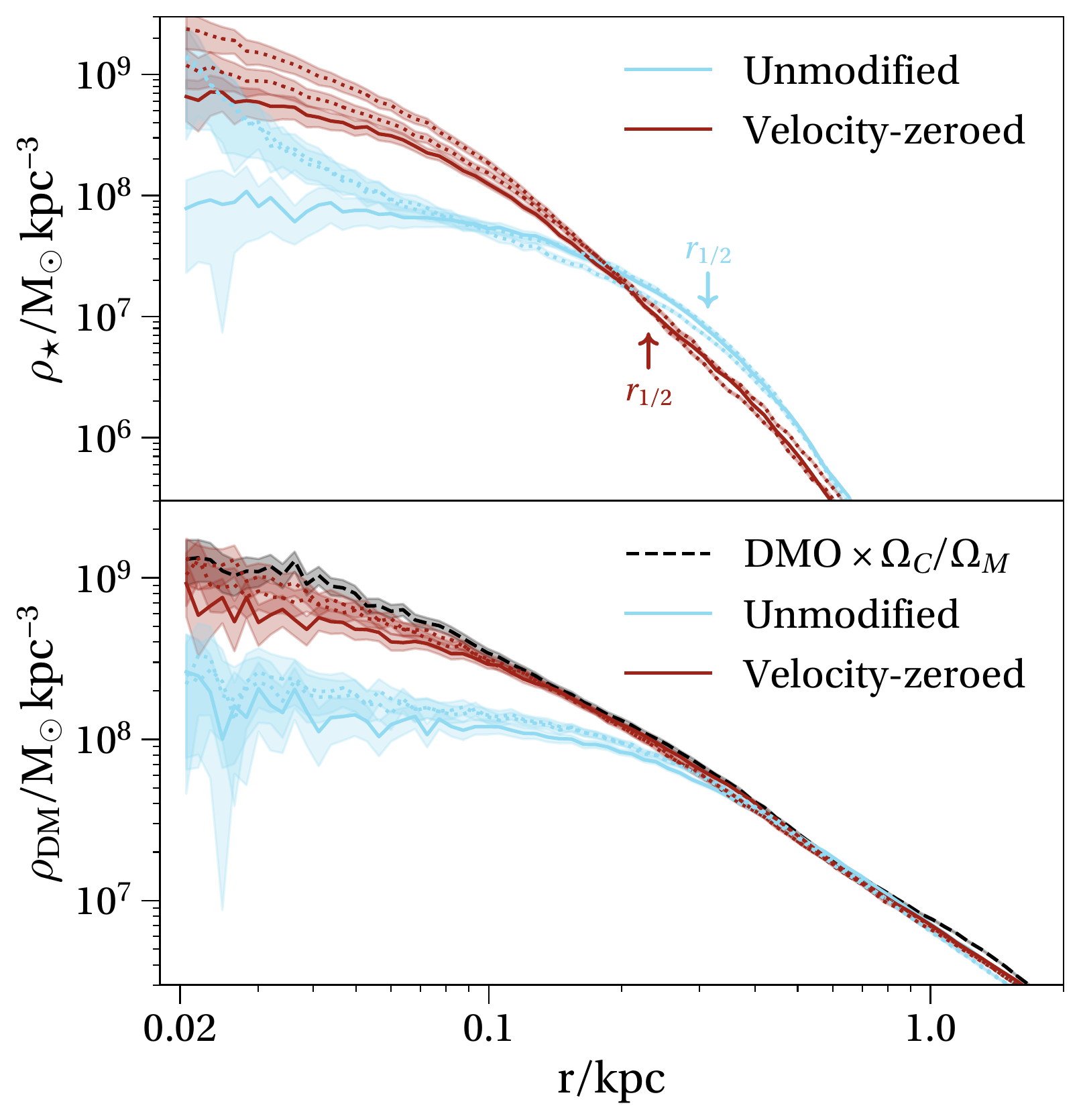}
  \caption{The stellar (top) and dark matter (bottom) spherically averaged density profiles at $z=0$. The colour scheme follows that of Figure~\ref{fig:SF-etc}, with an added black dashed line to show the density profile in the dark-matter-only simulation.  Shaded regions show r.m.s. uncertainty due to Poisson noise on the profiles. Star formation
    bursts in the unmodified runs (blue) lead to expansion of orbits, and so a dark matter core and a more diffuse stellar distribution. In the velocity-zeroed runs (red), the dark matter cusp survives and the stellar distribution is more concentrated.  Additionally, in two out of three runs with the unmodified initial conditions a dense central star cluster forms, constituting around $10\%$ of the total stellar mass. }\label{fig:stacked-den-profiles}
  \end{figure}

The final stellar mass is $4.8\pm 0.6 \times 10^6\,\Msol$ and $4.4 \pm 0.3 \times 10^6\,\Msol$ in the unmodified and velocity-zeroed cases respectively, showing that the early differences in star formation rate do not significantly  affect the final stellar mass. This can be understood by considering the overall supply of gas to the virial radius of the halo, which is near-independent of the initial conditions (see bottom panel of Figure~\ref{fig:SF-etc}). The key effect of diffusion is to keep the gas at lower densities for longer, hence delaying star formation without changing the overall supply of fuel in the circumgalactic medium. 

Despite the close agreement in stellar mass, the final distribution of stars is markedly different in the two cases. The top panel of Figure~\ref{fig:stacked-den-profiles} shows the spherically averaged stellar density between $20\,\pc$ and $2\,\kpc$. The minimum softening scale of $3\,\pc$ implies that the profiles are well resolved throughout this range; to increase precision we average the density over three timesteps, and plot the $r.m.s.$ error bands assuming uncorrelated Poisson noise.  The 3D stellar half-light radius (in the V band) is $300 \pm 20 \,\pc$ with the original initial conditions and $220 \pm 20\,\pc$ when velocity-zeroed; in other words, the final galaxy is more compact when it is allowed to form at rest with respect to the grid.  Notwithstanding this, in two out of three runs with the unmodified initial conditions we find a dense central stellar knot which has been formed during the extended tail in star formation discussed above. The knot is prominent in the stellar density profiles but does not impact significantly on the half-light radius because the total mass in this feature, in the two simulations where it is present, is only $4\times10^5 \Msol$ (less than a tenth of the total stellar mass). 

The difference between the half-light radii from different initial conditions is driven by the ability of stellar feedback to redistribute collisionless orbits \citep[for a review see][]{Pontzen14Review}. At $z=6$, the half-light radii are $108 \pm 40\,\pc$ and $115 \pm 20\,\pc$ for the original and velocity-zeroed initial conditions respectively: they are in agreement. The tail of star formation after this time injects supernova energy which, through gravitational potential fluctuations, expands the orbits of stars and dark matter. The expansion is thus much more pronounced in the galaxy forming from the unmodified initial conditions. The lower panel of Figure~\ref{fig:stacked-den-profiles} shows that an even stronger discrepancy between unmodified and velocity-zeroed simulations is observed for the dark matter density, with a significant core extending around $200\,\pc$ in the former case. The velocity-zeroed run, by contrast, retains a density profile almost as steep as that of the dark-matter-only run (black dashed line).

\begin{figure}
  \includegraphics[width=\columnwidth]{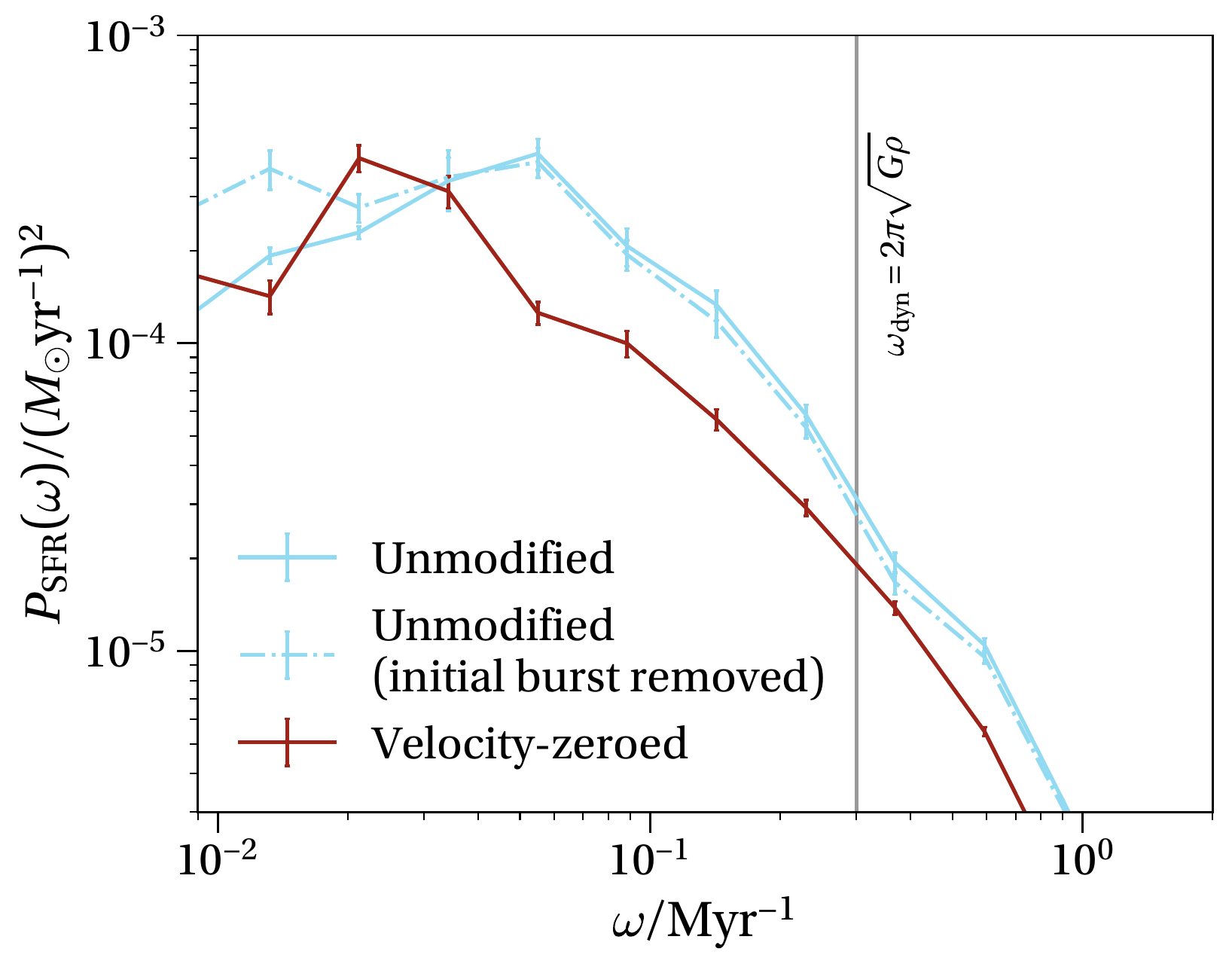}
  \caption{The power spectrum of the star formation rate, which gives a measure of time variability. The unmodified initial conditions (blue line) give rise to a burstier star formation rate, as quantified by the increased power compared to the velocity-zeroed simulation (red line). The dash-dotted line shows the star formation history of the unmodified initial conditions where the initial dramatic burst is excluded, showing that the enhanced burstiness is present throughout the star formation history.  Error bars show the r.m.s. uncertainty in the power spectrum estimates. Fluctuations near the dynamical time $\omega_{\mathrm{dyn}}$, shown with a vertical grey line,  are those which contribute to flattening dark matter cores and expanding stellar orbits. }\label{fig:sfr-variability}
  \end{figure}
 
  We next quantify the burstiness in the star formation, to help understand why the two initial conditions give rise to such different dynamical properties. Fluctuations in the gravitational potential sourced by bursty feedback can pump energy into collisionless orbits, provided changes occur over periods comparable to or shorter than the dynamical time $t_{\mathrm{dyn}}$  \citep{Pontzen12}. Fluctuations on longer timescales act adiabatically and thus cannot cause cumulative, long-term changes to stellar or dark matter distributions. To distinguish these regimes we require time resolution substantially better than $t_{\mathrm{dyn}} \equiv (G \rho)^{-1/2} \simeq 21\,\Myr$ in our dwarf at 100\,pc. Storing a sufficient number of snapshots (each measuring $9\,$GB) given that star formation extends over $>1\,\Gyr$ is out of reach for our study. Instead, we reconstruct the star formation history from the birth time of each star particle, using this as a proxy for the gas density and potential fluctuations.  

To compare the level of burstiness, we Fourier-transform the star formation history then produce a binned estimate of its modulus squared. This produces a power spectrum $P_{\mathrm{SFR}}(\omega)$ as a function of $\omega$, the wavenumber in units of $\Myr^{-1}$. We use 15 bins spaced equally in $\log_{10} \omega$ between $-2.6$ and $0.5$. The result is plotted in Figure~\ref{fig:sfr-variability}, along with $1\sigma$ uncertainty bars estimated from the variance within each bin, assuming Gaussian statistics. To reduce these errors and produce a cleaner measurement we average the power spectra from the three runs for each initial condition type. Larger power indicates greater star formation variability on the corresponding timescales $\Delta t = 2 \pi / \omega$. 

The results in Figure~\ref{fig:sfr-variability} show that the variability around the dynamical timescale $\omega_{\mathrm{dyn}} = 2 \pi/t_{\mathrm{dyn}}$ is larger in the unmodified initial conditions (blue line) compared to the more physical solution with velocity-zeroing (red line). It is therefore plausible that differences in the burstiness directly give rise to the difference in dark matter and stellar profiles between unmodified and velocity-zeroed cases.  Even if we exclude the initial large burst in the unmodified case (blue dash-dotted line), this remains true -- star formation proceeds in an excessively bursty mode at all times.  The likely cause is that, following a supernova explosion, numerical diffusion holds up the re-condensation of gas to star-forming densities. Just as with the initial burst, the delay causes gas to accumulate in the circumgalactic medium so that once condensation gets underway there is an unphysically catastrophic collapse and a new starburst. This cycle of alternating numerical diffusion and physical collapse ultimately gives rise to a galaxy with significantly altered dynamics. 

The actual burstiness of star formation in such dwarf galaxies at high redshift will be very hard to determine observationally, and within simulations is subject to a wide range of sub-grid and numerical uncertainties. Therefore, our results should not be taken to indicate that dark matter cores in dwarf spheroidal galaxies are necessarily unobtainable within $\Lambda$CDM \citep[although energy conservation sets a size limit to such cores; e.g.][]{Penbarrubia12,Read16}. The results serve only to illustrate that numerical diffusion should be suppressed to the maximum extent possible when studying the cusp-core question.

\section{Discussion and conclusions}\label{sec:disc}

We have studied how numerical diffusion coupled to streaming relative to a static mesh can unphysically soften gas filaments in the early Universe. The numerical diffusion scale can be orders of magnitude larger than the cell size of the simulation. In the context of the EDGE project to study faint dwarf galaxies, we have solved the problem by using genetic modification to zero the streaming velocities of the target region in the initial conditions. Genetic modification is an ideal tool for this manipulation, since it maintains the galaxy's accretion history and nearby large scale structure (Figure~\ref{fig:structure-panels}). While non-linear structure formation still generates a net grid-relative motion, it is suppressed by an order of magnitude and shifted to lower redshift (Figure~\ref{fig:DMO-history}).

Published EDGE results already use these `velocity-zeroed' initial conditions (Figure~\ref{fig:ICs}); in this paper we compared hydrodynamical simulations performed with and without the modification. We found that, without the corrective velocity-zeroing, numerical diffusion delays the onset of star formation by $320\,\Myr$. Once star formation finally does initiate in the unmodified case, a dramatic burst brings the total mass of stars into close agreement across all simulations; nonetheless, the overall properties of the final dwarf galaxy are substantially affected. In particular, the altered star formation history (Figure~\ref{fig:SF-etc}) leads to stronger potential fluctuations and hence an expanded stellar half-light radius and reduced central dark matter density (Figure~\ref{fig:stacked-den-profiles}).

The extent to which any given mesh-based simulation suffers from these kind of effects will depend on the cooling and feedback physics, the refinement criteria, the hydrodynamic solver and slope limiter, the accretion history of the galaxy and the nature of the observables under consideration. Advection errors are likely to be most problematic prior to reionisation, when there is little physical pressure smoothing in the intergalactic medium; and in the smallest objects, where the gravitational potential is insufficient to compress the gas in the presence of the numerical damping. These conditions bring to mind recent AMR-based simulations of reionisation by young galaxies \citep{XuWise16Renaissance,Trebitsch17,Katz18,2018RosdahlSPHINX,Trebitsch20}. Further investigation will be required to determine whether star formation efficiency and escape fractions in those works may have been affected by the effects of diffusion. \cite{Trebitsch17}, for example, have already noted that the time at which their first stars form, especially in their lowest mass halo, is dependent on resolution. This may be a reflection of the numerical diffusion's dependence on cell size $\Delta$, as seen in Eq.~\eqref{eq:diffusion-scale-estimator}, in agreement with our high-resolution test simulation (e.g. Figure~\ref{fig:SF-etc}).

\begin{figure}
  \includegraphics[width=1.0\columnwidth]{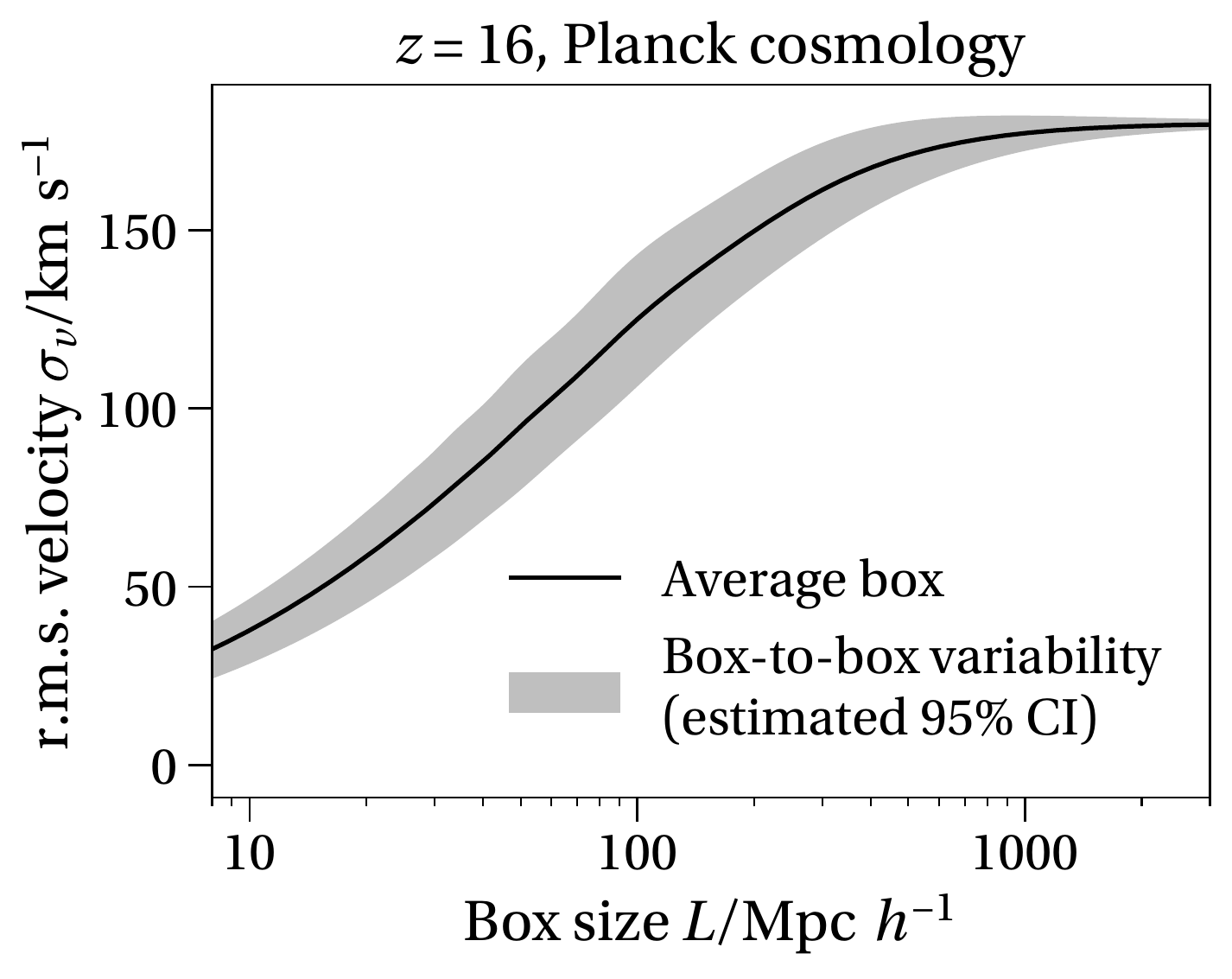}
  \caption{The expected linear r.m.s. velocity $\sigma_v$ relative to the grid within a single simulation box as a function of its comoving size $L$, evaluated at ${z=16}$ (the time at which stars start to form in our EDGE simulations).  Typical streaming velocities increase with the simulation volume, reaching a limit of $180\,\km\,\s^{-1}$ for a hypothetical simulation with $3\,\Gpc$ box size. Therefore, effects of grid-relative free-streaming are expected to be exacerbated in large cosmological volumes. 
  (Individual boxes will not have precisely this r.m.s. velocity, due to sample variance; the $95\,\%$ confidence interval from the estimated box-to-box scatter is shown as a grey band.)  } \label{fig:vel-rms}
  \end{figure}

Large peculiar velocities are common in cosmological simulations. Typical velocities also increase with box size, because they can be sourced by very long wavelength modes. The magnitude of the problem and its dependence on box size is illustrated in Figure~\ref{fig:vel-rms}, in which we plot the r.m.s. streaming velocities $\sigma_v$ as a function of box size $L$ at $z=16$ (the time at which stars start to form in the EDGE simulations), along with estimated $95\%$ box-to-box variability due to sample variance. The linear theory calculations to obtain this result are given in Appendix~\ref{sec:vel-calc}. For example, the r.m.s. velocity for our box size ($50\,\Mpc \simeq 33.9\,\Mpc/h$) is $79\,\kmps$, meaning that the halo examined in this work (if unmodified, streaming at $51\,\kmps$) actually suffers from diffusion somewhat {\it less} than a typical halo. As the box size increases and more long-wavelength modes start contributing, $\sigma_v$ reaches $125\,\kmps$ for $L=100\,\Mpc/h$; flattens to around $171\,\kmps$ for $L=500\,\Mpc/h$; and saturates at $180\,\kmps$ by $L=3\,\mathrm{Gpc}/h$. The box-to-box variations around this average are relatively small.

There are two consequences to this size dependence. As well as the obvious conclusion that larger AMR simulations may suffer proportionately more from diffusion effects, it also shows indirectly that velocities are always generated by some of the largest modes in the box; this is demonstrated more explicitly by defining and computing a velocity correlation length in Appendix~\ref{sec:vel-cor-length}. The practical conclusion is that velocity-zeroing should be achievable for objects of a wide range of masses, without affecting the merger history or local environment, just as we have seen for small galaxies in this paper. Provided the box size is very much larger than the Lagrangian region of the halo, we expect the modifications will remain strongly coherent across the entire patch, meaning that the dark matter accretion history and environment around the target object will be essentially unchanged.  On the other hand, the main limitation of velocity-zeroing is that it is only achievable for zoom simulations; we cannot suppress the streaming throughout the simulation domain without invoking an unphysical modification to the power spectrum (to remove all long-wavelength modes). 

The {\sc genetIC} code \citep{CodePaper} allows the user to perform the velocity-zeroing straight-forwardly. Similarly, any initial conditions generator that implements Hoffman-Ribak velocity constraints \citep[e.g.][]{HahnAbel11} can perform a similar manipulation, although we caution that at present we are not aware of codes (other than {\sc genetIC}) that correctly propagate constraints across different zoom levels. Given the simplicity of the operation, and the significance of its corrections, there is a strong case for incorporating the technique as standard for AMR zoom simulations.

\section*{Acknowledgements}

We gratefully acknowledge helpful comments on a draft from Julien Devriendt and John Wise, and from the anonymous referee. This project has received funding from the European Union's Horizon 2020 research and innovation programme under grant agreement No.  818085 GMGalaxies. AP was further supported by the Royal Society.   OA and MR acknowledge support from the Knut and Alice Wallenberg Foundation and the Royal Physiographic Society of Lund. OA is supported by the grant 2014-5791 from the Swedish Research Council. The authors acknowledge the use of the UCL Grace High Performance Computing Facility and associated support services. This work was performed in part using the DiRAC Data Intensive service at Leicester, operated by the University of Leicester IT Services, which is part of the STFC DiRAC HPC Facility (\url{www.dirac.ac.uk}). This work was also partially supported by the UCL Cosmoparticle Initiative. 

\section*{Data availability}
The data underlying this article will be shared on reasonable request to the corresponding author.

\bibliographystyle{mnras}
\bibliography{velshift}

\begin{thebibliography}{}
\makeatletter
\relax
\def\mn@urlcharsother{\let\do\@makeother \do\$\do\&\do\#\do\^\do\_\do\%\do\~}
\def\mn@doi{\begingroup\mn@urlcharsother \@ifnextchar [ {\mn@doi@}
  {\mn@doi@[]}}
\def\mn@doi@[#1]#2{\def\@tempa{#1}\ifx\@tempa\@empty \href
  {http://dx.doi.org/#2} {doi:#2}\else \href {http://dx.doi.org/#2} {#1}\fi
  \endgroup}
\def\mn@eprint#1#2{\mn@eprint@#1:#2::\@nil}
\def\mn@eprint@arXiv#1{\href {http://arxiv.org/abs/#1} {{\tt arXiv:#1}}}
\def\mn@eprint@dblp#1{\href {http://dblp.uni-trier.de/rec/bibtex/#1.xml}
  {dblp:#1}}
\def\mn@eprint@#1:#2:#3:#4\@nil{\def\@tempa {#1}\def\@tempb {#2}\def\@tempc
  {#3}\ifx \@tempc \@empty \let \@tempc \@tempb \let \@tempb \@tempa \fi \ifx
  \@tempb \@empty \def\@tempb {arXiv}\fi \@ifundefined
  {mn@eprint@\@tempb}{\@tempb:\@tempc}{\expandafter \expandafter \csname
  mn@eprint@\@tempb\endcsname \expandafter{\@tempc}}}

\bibitem[\protect\citeauthoryear{{Agertz} \& {Kravtsov}}{{Agertz} \&
  {Kravtsov}}{2015}]{AgertzKravtsov2015}
{Agertz} O.,  {Kravtsov} A.~V.,  2015, \mn@doi [\apj]
  {10.1088/0004-637X/804/1/18}, \href
  {http://adsabs.harvard.edu/abs/2015ApJ...804...18A} {804, 18}

\bibitem[\protect\citeauthoryear{{Agertz} et~al.,}{{Agertz}
  et~al.}{2007}]{Agertz07}
{Agertz} O.,  et~al., 2007, \mn@doi [\mnras]
  {10.1111/j.1365-2966.2007.12183.x}, \href
  {https://ui.adsabs.harvard.edu/abs/2007MNRAS.380..963A} {380, 963}

\bibitem[\protect\citeauthoryear{{Agertz}, {Kravtsov}, {Leitner}  \&
  {Gnedin}}{{Agertz} et~al.}{2013}]{Agertz13}
{Agertz} O.,  {Kravtsov} A.~V.,  {Leitner} S.~N.,   {Gnedin} N.~Y.,  2013,
  \mn@doi [\apj] {10.1088/0004-637X/770/1/25}, \href
  {https://ui.adsabs.harvard.edu/abs/2013ApJ...770...25A} {770, 25}

\bibitem[\protect\citeauthoryear{{Agertz} et~al.,}{{Agertz}
  et~al.}{2020}]{2019arXiv190402723A}
{Agertz} O.,  et~al., 2020, \mn@doi [\mnras] {10.1093/mnras/stz3053}, \href
  {https://ui.adsabs.harvard.edu/abs/2020MNRAS.491.1656A} {491, 1656}

\bibitem[\protect\citeauthoryear{{Bromm}}{{Bromm}}{2013}]{Bromm13FirstStars}
{Bromm} V.,  2013, \mn@doi [Reports on Progress in Physics]
  {10.1088/0034-4885/76/11/112901}, \href
  {https://ui.adsabs.harvard.edu/abs/2013RPPh...76k2901B} {76, 112901}

\bibitem[\protect\citeauthoryear{{Chabanier} et~al.,}{{Chabanier}
  et~al.}{2020}]{Chabanier20}
{Chabanier} S.,  et~al., 2020, Submitted to A\&A, \href
  {https://ui.adsabs.harvard.edu/abs/2020arXiv200704624C} {}

\bibitem[\protect\citeauthoryear{{Chisari} \& {Zaldarriaga}}{{Chisari} \&
  {Zaldarriaga}}{2011}]{Chisari11GaugeNewtSim}
{Chisari} N.~E.,  {Zaldarriaga} M.,  2011, \mn@doi [\prd]
  {10.1103/PhysRevD.83.123505}, \href
  {https://ui.adsabs.harvard.edu/abs/2011PhRvD..83l3505C} {83, 123505}

\bibitem[\protect\citeauthoryear{{Eisenstein} \& {Hut}}{{Eisenstein} \&
  {Hut}}{1998}]{1998ApJ...498..137E}
{Eisenstein} D.~J.,  {Hut} P.,  1998, \mn@doi [\apj] {10.1086/305535}, \href
  {https://ui.adsabs.harvard.edu/abs/1998ApJ...498..137E} {498, 137}

\bibitem[\protect\citeauthoryear{Few, Dobbs, Pettitt  \& Konstandin}{Few
  et~al.}{2016}]{Few16}
Few C.~G.,  Dobbs C.,  Pettitt A.,   Konstandin L.,  2016, \mn@doi [\mnras]
  {10.1093/mnras/stw1226}, 460, 4382

\bibitem[\protect\citeauthoryear{{Genel} et~al.,}{{Genel}
  et~al.}{2019}]{Genel19}
{Genel} S.,  et~al., 2019, \mn@doi [\apj] {10.3847/1538-4357/aaf4bb}, \href
  {https://ui.adsabs.harvard.edu/abs/2019ApJ...871...21G} {871, 21}

\bibitem[\protect\citeauthoryear{{Hahn} \& {Abel}}{{Hahn} \&
  {Abel}}{2011}]{HahnAbel11}
{Hahn} O.,  {Abel} T.,  2011, \mn@doi [\mnras]
  {10.1111/j.1365-2966.2011.18820.x}, \href
  {https://ui.adsabs.harvard.edu/\#abs/2011MNRAS.415.2101H} {415, 2101}

\bibitem[\protect\citeauthoryear{{Hoffman} \& {Ribak}}{{Hoffman} \&
  {Ribak}}{1991}]{HoffmanRibak91}
{Hoffman} Y.,  {Ribak} E.,  1991, \mn@doi [\apjl] {10.1086/186160}, \href
  {http://ukads.nottingham.ac.uk/abs/1991ApJ...380L...5H} {380, L5}

\bibitem[\protect\citeauthoryear{{Hopkins}}{{Hopkins}}{2015}]{Hopkins15}
{Hopkins} P.~F.,  2015, \mn@doi [\mnras] {10.1093/mnras/stv195}, \href
  {https://ui.adsabs.harvard.edu/abs/2015MNRAS.450...53H} {450, 53}

\bibitem[\protect\citeauthoryear{{Katz}, {Kimm}, {Haehnelt}, {Sijacki},
  {Rosdahl}  \& {Blaizot}}{{Katz} et~al.}{2018}]{Katz18}
{Katz} H.,  {Kimm} T.,  {Haehnelt} M.,  {Sijacki} D.,  {Rosdahl} J.,
  {Blaizot} J.,  2018, \mn@doi [\mnras] {10.1093/mnras/sty1225}, \href
  {https://ui.adsabs.harvard.edu/abs/2018MNRAS.478.4986K} {478, 4986}

\bibitem[\protect\citeauthoryear{{Keller}, {Wadsley}, {Wang}  \&
  {Kruijssen}}{{Keller} et~al.}{2019}]{Keller19}
{Keller} B.~W.,  {Wadsley} J.~W.,  {Wang} L.,   {Kruijssen} J.~M.~D.,  2019,
  \mn@doi [\mnras] {10.1093/mnras/sty2859}, \href
  {https://ui.adsabs.harvard.edu/abs/2019MNRAS.482.2244K} {482, 2244}

\bibitem[\protect\citeauthoryear{{Kim} et~al.,}{{Kim}
  et~al.}{2016}]{KimAgora16}
{Kim} J.-h.,  et~al., 2016, \mn@doi [\apj] {10.3847/1538-4357/833/2/202}, \href
  {https://ui.adsabs.harvard.edu/abs/2016ApJ...833..202K} {833, 202}

\bibitem[\protect\citeauthoryear{{Kretschmer}, {Agertz}  \&
  {Teyssier}}{{Kretschmer} et~al.}{2020}]{Kretschmer2020}
{Kretschmer} M.,  {Agertz} O.,   {Teyssier} R.,  2020, arXiv e-prints, \href
  {https://ui.adsabs.harvard.edu/abs/2020arXiv200303368K} {p. arXiv:2003.03368}

\bibitem[\protect\citeauthoryear{{Lewis}, {Challinor}  \& {Lasenby}}{{Lewis}
  et~al.}{2000}]{camb2000}
{Lewis} A.,  {Challinor} A.,   {Lasenby} A.,  2000, \mn@doi [\apj]
  {10.1086/309179}, \href
  {https://ui.adsabs.harvard.edu/abs/2000ApJ...538..473L} {538, 473}

\bibitem[\protect\citeauthoryear{{McNally}, {Lyra}  \& {Passy}}{{McNally}
  et~al.}{2012}]{McNally12}
{McNally} C.~P.,  {Lyra} W.,   {Passy} J.-C.,  2012, \mn@doi [\apjs]
  {10.1088/0067-0049/201/2/18}, \href
  {https://ui.adsabs.harvard.edu/abs/2012ApJS..201...18M} {201, 18}

\bibitem[\protect\citeauthoryear{{Mu{\~n}oz}, {Kratter}, {Springel}  \&
  {Hernquist}}{{Mu{\~n}oz} et~al.}{2014}]{Munoz14}
{Mu{\~n}oz} D.~J.,  {Kratter} K.,  {Springel} V.,   {Hernquist} L.,  2014,
  \mn@doi [\mnras] {10.1093/mnras/stu1918}, \href
  {https://ui.adsabs.harvard.edu/abs/2014MNRAS.445.3475M} {445, 3475}

\bibitem[\protect\citeauthoryear{{Pe{\~n}arrubia}, {Pontzen}, {Walker}  \&
  {Koposov}}{{Pe{\~n}arrubia} et~al.}{2012}]{Penbarrubia12}
{Pe{\~n}arrubia} J.,  {Pontzen} A.,  {Walker} M.~G.,   {Koposov} S.~E.,  2012,
  \mn@doi [\apjl] {10.1088/2041-8205/759/2/L42}, \href
  {https://ui.adsabs.harvard.edu/abs/2012ApJ...759L..42P} {759, L42}

\bibitem[\protect\citeauthoryear{{Planck Collaboration} et~al.,}{{Planck
  Collaboration} et~al.}{2014}]{2014A&A...571A..16P}
{Planck Collaboration} et~al., 2014, \mn@doi [\aap]
  {10.1051/0004-6361/201321591}, \href
  {https://ui.adsabs.harvard.edu/abs/2014A&A...571A..16P} {571, A16}

\bibitem[\protect\citeauthoryear{{Pontzen} \& {Governato}}{{Pontzen} \&
  {Governato}}{2012}]{Pontzen12}
{Pontzen} A.,  {Governato} F.,  2012, \mn@doi [\mnras]
  {10.1111/j.1365-2966.2012.20571.x}, \href
  {https://ui.adsabs.harvard.edu/abs/2012MNRAS.421.3464P} {421, 3464}

\bibitem[\protect\citeauthoryear{{Pontzen} \& {Governato}}{{Pontzen} \&
  {Governato}}{2014}]{Pontzen14Review}
{Pontzen} A.,  {Governato} F.,  2014, \mn@doi [\nat] {10.1038/nature12953},
  \href {https://ui.adsabs.harvard.edu/abs/2014Natur.506..171P} {506, 171}

\bibitem[\protect\citeauthoryear{{Pontzen} \& {Tremmel}}{{Pontzen} \&
  {Tremmel}}{2018}]{2018ApJS..237...23P}
{Pontzen} A.,  {Tremmel} M.,  2018, \mn@doi [\apjs] {10.3847/1538-4365/aac832},
  \href {https://ui.adsabs.harvard.edu/abs/2018ApJS..237...23P} {237, 23}

\bibitem[\protect\citeauthoryear{{Pontzen}, {Ro{\v{s}}kar}, {Stinson}  \&
  {Woods}}{{Pontzen} et~al.}{2013}]{2013ascl.soft05002P}
{Pontzen} A.,  {Ro{\v{s}}kar} R.,  {Stinson} G.,   {Woods} R.,  2013, {pynbody:
  N-Body/SPH analysis for python} (\mn@eprint {ascl} {1305.002})

\bibitem[\protect\citeauthoryear{{Read}, {Agertz}  \& {Collins}}{{Read}
  et~al.}{2016}]{Read16}
{Read} J.~I.,  {Agertz} O.,   {Collins} M.~L.~M.,  2016, \mn@doi [\mnras]
  {10.1093/mnras/stw713}, \href
  {https://ui.adsabs.harvard.edu/abs/2016MNRAS.459.2573R} {459, 2573}

\bibitem[\protect\citeauthoryear{{Rey} \& {Pontzen}}{{Rey} \&
  {Pontzen}}{2018}]{Rey18}
{Rey} M.~P.,  {Pontzen} A.,  2018, \mn@doi [\mnras] {10.1093/mnras/stx2744},
  \href {https://ui.adsabs.harvard.edu/abs/2018MNRAS.474...45R} {474, 45}

\bibitem[\protect\citeauthoryear{{Rey}, {Pontzen}, {Agertz}, {Orkney}, {Read},
  {Saintonge}  \& {Pedersen}}{{Rey} et~al.}{2019}]{Rey19}
{Rey} M.~P.,  {Pontzen} A.,  {Agertz} O.,  {Orkney} M. D.~A.,  {Read} J.~I.,
  {Saintonge} A.,   {Pedersen} C.,  2019, \mn@doi [\apjl]
  {10.3847/2041-8213/ab53dd}, \href
  {https://ui.adsabs.harvard.edu/abs/2019ApJ...886L...3R} {886, L3}

\bibitem[\protect\citeauthoryear{{Rey}, {Pontzen}, {Agertz}, {Orkney}, {Read}
  \& {Rosdahl}}{{Rey} et~al.}{2020}]{Rey20}
{Rey} M.~P.,  {Pontzen} A.,  {Agertz} O.,  {Orkney} M. D.~A.,  {Read} J.~I.,
  {Rosdahl} J.,  2020, arXiv e-prints, \href
  {https://ui.adsabs.harvard.edu/abs/2020arXiv200409530R} {p. arXiv:2004.09530}

\bibitem[\protect\citeauthoryear{{Robertson}, {Kravtsov}, {Gnedin}, {Abel}  \&
  {Rudd}}{{Robertson} et~al.}{2010}]{Robertson10}
{Robertson} B.~E.,  {Kravtsov} A.~V.,  {Gnedin} N.~Y.,  {Abel} T.,   {Rudd}
  D.~H.,  2010, \mn@doi [\mnras] {10.1111/j.1365-2966.2009.15823.x}, \href
  {https://ui.adsabs.harvard.edu/abs/2010MNRAS.401.2463R} {401, 2463}

\bibitem[\protect\citeauthoryear{{Rosdahl} et~al.,}{{Rosdahl}
  et~al.}{2018}]{2018RosdahlSPHINX}
{Rosdahl} J.,  et~al., 2018, \mn@doi [\mnras] {10.1093/mnras/sty1655}, \href
  {https://ui.adsabs.harvard.edu/abs/2018MNRAS.479..994R} {479, 994}

\bibitem[\protect\citeauthoryear{Roth, Pontzen  \& Peiris}{Roth
  et~al.}{2016}]{Roth:2015wha}
Roth N.,  Pontzen A.,   Peiris H.~V.,  2016, \mn@doi [MNRAS]
  {10.1093/mnras/stv2375}, 455, 974

\bibitem[\protect\citeauthoryear{{Scannapieco} et~al.,}{{Scannapieco}
  et~al.}{2012}]{Aquila12}
{Scannapieco} C.,  et~al., 2012, \mn@doi [\mnras]
  {10.1111/j.1365-2966.2012.20993.x}, \href
  {https://ui.adsabs.harvard.edu/abs/2012MNRAS.423.1726S} {423, 1726}

\bibitem[\protect\citeauthoryear{{Somerville} \& {Dav{\'e}}}{{Somerville} \&
  {Dav{\'e}}}{2015}]{SomervilleDave15}
{Somerville} R.~S.,  {Dav{\'e}} R.,  2015, \mn@doi [\araa]
  {10.1146/annurev-astro-082812-140951}, \href
  {https://ui.adsabs.harvard.edu/abs/2015ARA&A..53...51S} {53, 51}

\bibitem[\protect\citeauthoryear{{Springel}}{{Springel}}{2010a}]{Springel10SPHReview}
{Springel} V.,  2010a, \mn@doi [\araa] {10.1146/annurev-astro-081309-130914},
  \href {https://ui.adsabs.harvard.edu/abs/2010ARA&A..48..391S} {48, 391}

\bibitem[\protect\citeauthoryear{{Springel}}{{Springel}}{2010b}]{Arepo10}
{Springel} V.,  2010b, \mn@doi [\mnras] {10.1111/j.1365-2966.2009.15715.x},
  \href {https://ui.adsabs.harvard.edu/abs/2010MNRAS.401..791S} {401, 791}

\bibitem[\protect\citeauthoryear{{Stopyra}, {Pontzen}, {Peiris}, {Roth}  \&
  {Rey}}{{Stopyra} et~al.}{2020}]{CodePaper}
{Stopyra} S.,  {Pontzen} A.,  {Peiris} H.,  {Roth} N.,   {Rey} M.,  2020,
  Submitted to ApJS, \href
  {https://ui.adsabs.harvard.edu/abs/2020arXiv200601841S} {p. arXiv:2006.01841}

\bibitem[\protect\citeauthoryear{{Teyssier}}{{Teyssier}}{2002}]{Teyssier02Ramses}
{Teyssier} R.,  2002, \mn@doi [\aap] {10.1051/0004-6361:20011817}, \href
  {https://ui.adsabs.harvard.edu/abs/2002A&A...385..337T} {385, 337}

\bibitem[\protect\citeauthoryear{{Teyssier}}{{Teyssier}}{2015}]{Teyssier15AMRReview}
{Teyssier} R.,  2015, \mn@doi [\araa] {10.1146/annurev-astro-082214-122309},
  \href {https://ui.adsabs.harvard.edu/abs/2015ARA&A..53..325T} {53, 325}

\bibitem[\protect\citeauthoryear{{Toro}}{{Toro}}{2009}]{Toro99}
{Toro} E.~F.,  2009, {Riemann Solvers and Numerical Methods for Fluid
  Dynamics}.
Springer Berlin Heidelberg, \url
  {https://books.google.co.uk/books?id=SqEjX0um8o0C}

\bibitem[\protect\citeauthoryear{{Toro}, {Spruce}  \& {Speares}}{{Toro}
  et~al.}{1994}]{1994ShWav...4...25T}
{Toro} E.~F.,  {Spruce} M.,   {Speares} W.,  1994, \mn@doi [Shock Waves]
  {10.1007/BF01414629}, \href
  {https://ui.adsabs.harvard.edu/abs/1994ShWav...4...25T} {4, 25}

\bibitem[\protect\citeauthoryear{{Trebitsch}, {Blaizot}, {Rosdahl}, {Devriendt}
   \& {Slyz}}{{Trebitsch} et~al.}{2017}]{Trebitsch17}
{Trebitsch} M.,  {Blaizot} J.,  {Rosdahl} J.,  {Devriendt} J.,   {Slyz} A.,
  2017, \mn@doi [\mnras] {10.1093/mnras/stx1060}, \href
  {https://ui.adsabs.harvard.edu/abs/2017MNRAS.470..224T} {470, 224}

\bibitem[\protect\citeauthoryear{{Trebitsch} et~al.,}{{Trebitsch}
  et~al.}{2020}]{Trebitsch20}
{Trebitsch} M.,  et~al., 2020, Submitted to A\&A, \href
  {https://ui.adsabs.harvard.edu/abs/2020arXiv200204045T} {p. arXiv:2002.04045}

\bibitem[\protect\citeauthoryear{{Tseliakhovich} \& {Hirata}}{{Tseliakhovich}
  \& {Hirata}}{2010}]{TseliakhovichHirata10}
{Tseliakhovich} D.,  {Hirata} C.,  2010, \mn@doi [\prd]
  {10.1103/PhysRevD.82.083520}, \href
  {https://ui.adsabs.harvard.edu/abs/2010PhRvD..82h3520T} {82, 083520}

\bibitem[\protect\citeauthoryear{{Weinberger}, {Springel}  \&
  {Pakmor}}{{Weinberger} et~al.}{2020}]{Arepo19}
{Weinberger} R.,  {Springel} V.,   {Pakmor} R.,  2020, \mn@doi [\apjs]
  {10.3847/1538-4365/ab908c}, \href
  {https://ui.adsabs.harvard.edu/abs/2020ApJS..248...32W} {248, 32}

\bibitem[\protect\citeauthoryear{{Xu}, {Wise}, {Norman}, {Ahn}  \&
  {O'Shea}}{{Xu} et~al.}{2016}]{XuWise16Renaissance}
{Xu} H.,  {Wise} J.~H.,  {Norman} M.~L.,  {Ahn} K.,   {O'Shea} B.~W.,  2016,
  \mn@doi [\apj] {10.3847/1538-4357/833/1/84}, \href
  {https://ui.adsabs.harvard.edu/abs/2016ApJ...833...84X} {833, 84}

\bibitem[\protect\citeauthoryear{{van de Weygaert} \& {Bertschinger}}{{van de
  Weygaert} \& {Bertschinger}}{1996}]{vanDeWeygaert96}
{van de Weygaert} R.,  {Bertschinger} E.,  1996, \mn@doi [\mnras]
  {10.1093/mnras/281.1.84}, \href
  {https://ui.adsabs.harvard.edu/abs/1996MNRAS.281...84V} {281, 84}

\makeatother
\end{thebibliography}

\appendix

\section{Expectations for numerical diffusion}\label{sec:advection-test}

In order to confirm that differences in our simulations are accounted for by numerical diffusion, in Section~\ref{sec:results} we estimated the expected smoothing using Eq.~\eqref{eq:diffusion-scale-estimator}. In this Appendix, we provide the derivation of that relationship. We study an idealised 1D advection problem, where an initial density distribution $\rho(x,t=0)$ moves at constant speed $v$ relative to the grid.  We will first provide some analytic insights, and then describe a numerical test which leads to the desired estimate. 

The diffusion-advection equation takes the form 
\begin{equation}
\frac{\partial \rho}{\partial t} + v \frac{\partial \rho}{\partial x} = \nu_{\mathrm{num}} \frac{\partial^2 \rho}{\partial x^2}\,.\label{eq:diff-adv-1st}
\end{equation}
The term on the right-hand side is generated by inaccuracies in the first-order gradient estimates, and $\nu_{\mathrm{num}}$ may be  obtained by replacing derivatives in the ideal advection problem with Taylor-expansions of the discrete numerical gradient estimates \citep[e.g.][]{Toro99}.  
Provided that the timestep is sufficiently small compared with the Courant-Friedrichs-Lewy (CFL) time, the numerical diffusion coefficient for the first-order upwind scheme reduces to $\nu_{\mathrm{num}} \simeq 2|v|\Delta$. By Fourier transforming Eq.~\eqref{eq:diff-adv-1st} in space, one obtains the solution 
\begin{equation}
\tilde\rho(k,t) = \tilde\rho(k,0)\, \ee^{-ikvt}\, \ee^{ -2k^2 |v|t\Delta}\,,\label{eq:tilderho}
\end{equation}
where $\tilde{\rho}(k,t)$ is the Fourier transform of $\rho(x,t)$.
The first exponential in Eq.~\eqref{eq:tilderho} represents the physical advection, while the second represents a Gaussian filter, suppressing high$-k$ modes due to the effect of numerical diffusion. Note that $|v|t=L$, the total distance travelled across the grid at time $t$. When the inverse Fourier transform is applied to Eq.~\eqref{eq:tilderho}, one obtains the displaced original density distribution convolved with a Gaussian of width   
\begin{equation}
  \sigma_{\mathrm{1st}} \approx L^{1/2} \Delta^{1/2}\,.\label{eq:first-order-diffusion}
\end{equation}
However, our configuration of {\sc Ramses} for the EDGE suite does not use a purely first-order scheme; rather it uses a hybrid between first-order and second-order slope estimation known as the MinMod scheme. In regions of the intergalactic medium that are relatively smooth, the scheme should come close to the performance of a pure second-order solver. In such a case, Taylor expansion of the numerical derivatives reveals the errors are third order,
\begin{equation}
  \frac{\partial \rho}{\partial t} + v \frac{\partial \rho}{\partial x} = \alpha_{\mathrm{num}} \frac{\partial^3 \rho}{\partial x^3}\,.
\end{equation}
where $\alpha_{\mathrm{num}} = \beta |v| \Delta^2$, with a scheme-dependent numerical prefactor $\beta$. We have again assumed that the timestep is much smaller than the CFL time. Instead of a pure Gaussian spread, the Fourier-space solution now involves a dispersion:
\begin{equation}
  \tilde\rho(k,t) = \tilde\rho(k,0)\, \ee^{-ikvt}\, \ee^{ i \beta k^3 vt \Delta^2 }\,.\label{eq:second-order-f-space}
\end{equation}
This is poorly behaved as $k$ becomes large, giving rise to oscillatory downstream instabilities for pure second-order schemes. In practice, as stated above, second-order schemes are regularised using a slope limiter which cannot be represented in a linear analysis. All one can say from equation~\eqref{eq:second-order-f-space} is that there is a critical wavenumber $k_{\mathrm{crit}}$ at which errors must become significant, $\beta \Delta^2 k_{\mathrm{crit}}^3 |v| t \approx 1$. From this, we estimate the scaling of the diffusion in second order schemes:
\begin{equation}
  \sigma_{\mathrm{2nd}} \propto k_{\mathrm{crit}}^{-1} \propto L^{1/3} \Delta^{2/3}\,.\label{eq:second-order-diffusion}
\end{equation}
The steeper scaling with cell size $\Delta$ compared to the first order case~\eqref{eq:first-order-diffusion} implies that the effective numerical diffusion can be suppressed more efficiently with increasing resolution. 

To test whether this argument holds, we used {\sc Ramses} to perform 1D simulations; the purpose is to estimate the width of numerical diffusion as an improvement on Equation~\eqref{eq:second-order-diffusion}.  Specifically, we initialise a periodic box with $2^8$ cells at uniform density and temperature. We then introduce by hand a $50\%$ enhancement in density to a single cell at the centre of the box.  The perturbation is chosen to be in pressure equilibrium with its surroundings and moreover we do not include any gravity or cooling in our simulation, so that the problem remains one of scale-free advection.  To measure the diffusion width as a function of time we impart a uniform velocity $v$ to the entire grid so that we directly test the accuracy of the advection.   We ensure our time steps are much smaller than the CFL time.

For each snapshot we fit a Gaussian with width $\sigma$; the fit remains qualitatively good even at late times, although analysis in Fourier space shows that the high-$k$ behaviour departs from an exact Gaussian (as expected given the higher-order solver). The width $\sigma$ is fit extremely well by the relation 
\begin{equation}
  \sigma \approx 0.9 L^{0.29} \Delta^{0.71}\,,\label{eq:diffusion-scale-estimator-in-appendix}
\end{equation}
which is very close to the expected trend, Eq.~\eqref{eq:second-order-diffusion}. 

Our tests were run with the HLLC solver and MinMod slope limiter, to match the choices in the EDGE cosmological suite. We caution, however, that the estimated diffusion scales do not take into account the intrinsic effect of mesh adaptation, nor the related fact that the structure being advected forms from initial small fluctuations rather than being passively advected. Nonetheless, these scales do permit a helpful qualitative interpretation of the difference between unmodified and velocity-zeroed simulations. In Section~\ref{sec:results}, we showed that Eq.~\eqref{eq:diffusion-scale-estimator-in-appendix}  gives consistent estimates with the scales on which smoothing is seen in our simulations. 

\section{Calculating statistical properties of the velocity field in cosmological initial conditions}
\subsection{Typical grid-relative velocities as a function of box size}\label{sec:vel-calc}

In concluding the paper, we drew attention to the steep rise in r.m.s. velocity $\sigma_v$ as a function of computational box size (Figure~\ref{fig:vel-rms}). This Appendix provides a derivation of that trend. 

To compute the typical velocity with which galaxies will be streaming relative to the grid, we assume that large-scale cosmic flows are dominated by the linear field, at least at the high redshifts of interest.  We assume that the real-space overdensity field $\delta(\vec{x},z)$ (at a given comoving position $\vec{x}$ and redshift $z$) is linked to the Fourier-space overdensities $\delta(\vec{k},z)$ with the convention 
\begin{equation}
  \delta(\vec{x},z) = \frac{1}{(2 \pi)^3} \int \dd^3 \vec{k} \, \ee^{i \vec{k} \cdot \vec{x}} \delta(\vec{k},z)\,.
\end{equation}
We can then define the power spectrum of density fluctuations $P(k,z)$ via 
\begin{equation}
\langle \delta(\vec{k},z) \delta(\vec{k}',z) \rangle = (2 \pi)^3 \delta_D(\vec{k} + \vec{k}') P(k,z)\,,
\end{equation}
where $\delta_D$ denotes the Dirac delta function, and $k = |\vec{k}|$. The one-point r.m.s. density fluctuations $\sigma_{\delta}$ are given by 
\begin{equation}
\sigma_{\delta}^2 = \langle | \delta(\vec{x}) | ^2 \rangle = \frac{1}{2 \pi^2} \int k^2 P(k,z) \, \dd k\,.\label{eq:density-variance}
\end{equation}
Typically, values of $\sigma$ are only quoted after a filter has been applied to the field to smooth small-scale fluctuations. However, for velocity (as opposed to density) fluctuations, the small-scale fluctuations are a subdominant contribution as we shall see, and no smoothing will be required to obtain a converged result. 

We wish to find the equivalent expression to~\eqref{eq:density-variance} but for r.m.s. fluctuations in velocity.
As we will consider box sizes of several Gpc, relativistic gauge freedom must be considered. While numerical simulations by construction compute density fields in synchronous gauge, their velocity fields reside in conformal Newtonian gauge \citep[e.g.][]{Chisari11GaugeNewtSim}. During matter domination, when potentials are constant, the physical velocities and overdensities are then related by 
\begin{equation}
  v_i(\vec{k},z) = \frac{-iH(z) k_i}{(1+z)k^2}\, \delta(\vec{k},z)\,,\label{eq:velocity-from-overden}
\end{equation}
where $v_i(\vec{k}, z)$ is the Fourier transform of the velocity field components for wavenumber $\vec{k}$, $\delta(\vec{k},z)$ is the corresponding overdensity, $z$ is the redshift for which we are computing the fields and $H(z)$ is the Hubble parameter. 
Consequently, the expected variance on the velocity within a cube of comoving size $L$ is given by 
\begin{equation}
\langle \sigma_v^2 \rangle = \langle |\vec{v}(\vec{x})|^2 \rangle = \frac{H(z)^2}{2 \pi^2 (1+z)^2} \int_{2\pi / L}^{\infty} P(k,z) \,\dd k\,.\label{eq:velocity-variance}
\end{equation}
In the physical Universe, the limits of the integral are between $k=0$ and $k=\infty$ but in the numerical Universe there are cutoffs at both ends, provided by the particle spacing (high $k$) and box size (low $k$); well-resolved simulations have sufficient resolution to ignore the former, but  the latter must be given by $2 \pi / L$. The missing factor $k^2$ compared to the density case~\eqref{eq:density-variance} means that the velocity variance integral~\eqref{eq:velocity-variance} is typically dominated by low-$k$ modes. Therefore in our estimate we ignore any high-$k$ cut-off or filtering scale and integrate from $k_{\mathrm{min}} = 2 \pi / L$ where $L$ is the box size. We evaluate $P(k)$ for our adopted $\Lambda$CDM cosmology (see Section~\ref{sec:sims}) using \texttt{CAMB} \citep{camb2000}. This gives rise to the function $\sigma_v(L)$ which is plotted in Figure~\ref{fig:vel-rms}.

The steep rise in $\sigma_v$ as a function of $L$ up to several hundred ${\rm Mpc}/h$ comoving implies that flows will be sourced by the fundamental or near-fundamental modes in cosmological simulations. In turn this suggests that there may be a significant box-to-box variability in $\sigma_v$ since the sampling of such modes is sparse, exacerbating sample variance. To quantify the scatter, we must drop the assumption of the continuum limit and instead write 
\begin{equation}
\delta(\vec x) = \sum_{\vec k} \ee^{i \vec k \cdot \vec x} \delta_{\vec k}\,,\label{eq:discrete-ft}
\end{equation}
where the sum is over all modes between the fundamental and Nyquist frequencies.  We do not consider the effect of the Nyquist cutoff here, since for practical simulation geometries it has little effect on the magnitude of velocity fluctuations, for the reasons given immediately below Eq.~\eqref{eq:velocity-variance}. The discrete $\delta_{\vec k}$ obey Gaussian statistics with 
\begin{align}
\langle \delta_{\vec k}(z) \delta_{\vec k'}(z)^* \rangle & = \frac{1}{L^3} \delta_{\vec k\vec k'} P(k,z)\,.\label{eq:discrete-power}
\end{align}
where the Kronecker delta $\delta_{\vec k \vec k '}=1$ if $\vec{k}=\vec{k}'$ and 0 otherwise; and the wavenumber $k$ is defined by $k\equiv|\vec{k}|$. Starting from Eq.~\eqref{eq:discrete-ft} and applying Wick's theorem alongside Eq.~\eqref{eq:discrete-power}, one may now show that 
\begin{equation}
\langle|\delta(x)^4|\rangle - \langle|\delta(x)^2|\rangle^2 = 2 \sum_{\vec{k}} \frac{P(k,z)^2}{L^6}\,.
\end{equation}
When dealing with the velocity magnitude instead of the overdensity we pick up four factors of the conversion ratio \eqref{eq:velocity-from-overden}
and so, returning to the continuum approximation, we can quantify the expected box-to-box scatter in $\sigma_v$ as 
\begin{equation}\label{eq:velocity-4th-moment}
\langle (\sigma_v^2)^2 \rangle - \langle \sigma_v^2 \rangle^2 = \frac{H(z)^4}{ \pi^2 L^3 (1+z)^4 } \int_{2 \pi / L}^{\infty} \dd k\, k^{-2} P(k,z)^2\,.
\end{equation}
This final expression retains two dependences on $L$. The dependence in the integration limits can be traced to the cut-off in power above the box scale; the pre-factor $L^{-3}$, on the other hand, can be traced to the density of sampling in Fourier space as a function of box size. To turn the expression into the confidence interval shown in Figure~\ref{fig:vel-rms}, we assume that the distribution in $\sigma_v$ is Gaussian, with second and fourth moments as expressed in equations~\eqref{eq:velocity-variance} and~\eqref{eq:velocity-4th-moment}. This does not describe the precise distribution of velocities, which is in fact a generalised chi-squared distribution in $\sigma_v^2$. Nonetheless we verified by comparison to $1\,000$ realisations of the velocity field with a box size of $L=60\,\Mpc/h$ that the Gaussian limit constitutes a good approximation, with the main difference being that the true distribution has a slight skew towards a high $\sigma_v$ tail. 

\begin{figure}
\includegraphics[width=\columnwidth]{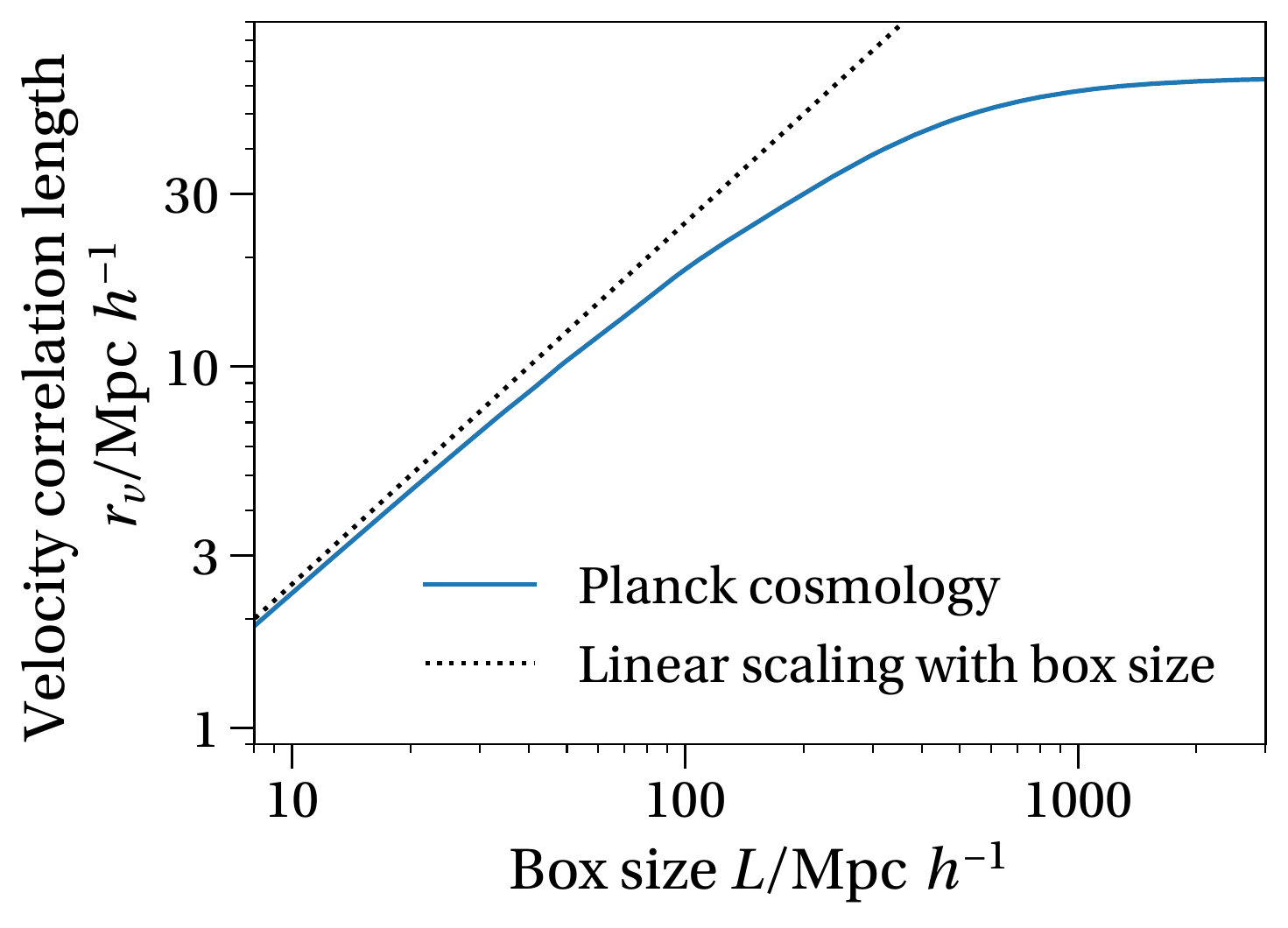}
\caption{Representative comoving distance $r_v$ over which the velocity remains correlated, as a function of simulation comoving box size $L$. For small box sizes, $r_v$ scales proportionally to $L$ because the velocity power is dominated by waves near the fundamental mode. (The dotted line shows a linear scaling for comparison.) For $L>100\,\Mpc/h$, the correlation length begins to flatten until it reaches its cosmological value $r_v \simeq 63\,\Mpc/h$. }\label{fig:vel-correlation-length}
\end{figure}

\subsection{Velocity correlation length as a function of box size}\label{sec:vel-cor-length}

Figure~\ref{fig:ICs} shows that, in our specific case, the modification is coherent over a region far larger than the Lagrangian patch of our simulated galaxy. This in turn ensures that its halo accretion history and environment are unchanged by the velocity-zeroing operation. In the conclusions (Section~\ref{sec:disc}) we claimed that such coherence can be arranged even for larger halos, by sufficiently increasing the simulation box size $L$. We now provide evidence for that claim, by calculating the distance over which the initial velocity field is correlated. 

Consider the velocity cross-correlation between two points separated by a distance $r$, which we define by $\xi_v(r) \equiv \langle \vec{v}(\vec{0}) \cdot \vec{v}(\vec{r}) \rangle$; by statistical homogeneity we take the first point to be the origin, and by statistical isotropy the displacement $\vec{r}$ can be chosen to lie along any direction. Following the same arguments leading to Eq.~\eqref{eq:velocity-variance}, we obtain
\begin{equation}
  \xi_v(r)  = \frac{H(z)^2}{2 \pi^2 (1+z)^2} \int_{2\pi / L}^{\infty} P(k,z) \frac{\sin\,kr}{kr}\,\dd k\,.\label{eq:velocity-covariance}
\end{equation}
The correlation function $\xi_v$ correctly boils down to Eq.~\eqref{eq:velocity-variance} in the limit $r \to 0$. As $r \to \infty$, $\xi_v(r)$ oscillates and dies away to zero. 

We now define the velocity correlation length $r_v$ (in comoving units) to be the value of $r$ at which the correlation has declined by a factor of $4$, i.e. $\xi_v(r_v) \equiv \xi_v(0)/4$. The factor $4$ is somewhat arbitrary; adopting another value rescales the quantitative values for $r_v$, but does not lead to qualitatively different conclusions. Note that $r_v$ is independent of redshift in the linear approximation.

We evaluate $r_v(L)$ numerically using the $\Lambda$CDM cosmological power spectrum $P(k)$ as above, and plot the result in Figure~\ref{fig:vel-correlation-length}. For $L<100\,\Mpc/h$, $r_v$ increases linearly with $L$ because the velocities are strongly dominated by the near-fundamental modes. The velocity modifications are coherent over a significant fraction of the box, which is consistent with the visual impression given by Figure~\ref{fig:ICs}. At larger $L$, as the overall r.m.s. velocity begins to saturate (Figure~\ref{fig:vel-rms}), the correlation length grows more slowly. 
 For very large boxes ($L = 3\,\Gpc/h$) patches as large as $r_v \simeq 63 \,\Mpc/h$ are well-correlated in velocity. There is no benefit in further increasing $L$ since $r_v$ is now fully determined by the cosmological model rather than the finite numerical domain. 

Given the large values of $r_v$ attainable, we conclude that for zoom galaxy formation simulations, it should be possible to make velocity modifications without altering the accretion history and local large scale structure in any significant way. For example, a galaxy of mass $10^{12}\,\Msol/h$ collapses from a region of comoving radius $\simeq 1.4\,\Mpc/h$; for a cluster of mass $10^{14}\,\Msol/h$, the corresponding value is $\simeq 6.5\,\Mpc/h$, still a factor $10$ lower than the limiting value of $r_v$.

\end{document}